\documentclass[11pt]{article}

\usepackage[T1]{fontenc}
\usepackage[utf8]{inputenc} 
\usepackage{lmodern}
\usepackage{microtype}
\usepackage{geometry}
\usepackage{graphicx}
\graphicspath{{figures/}}
\usepackage{booktabs}
\usepackage{multirow}
\usepackage{amsmath, amssymb}
\usepackage{authblk}
\usepackage{hyperref}
\hypersetup{colorlinks=true,linkcolor=blue,citecolor=blue,urlcolor=blue}

\title{Synthetic Data Generation for Classifying Electrophysiological and Morpho-Electrophysiological Neurons from Mouse Visual Cortex}

\author[1]{Xavier Vasques}
\author[1,2,3]{Laura Cif}

\affil[1]{Institut du Neurone, Montferriez sur lez, France}
\affil[2]{Département de Neurochirurgie, Unité des Pathologies Cérébrales Résistantes, Hôpital Gui de Chauliac, Centre Hospitalier Universitaire de Montpellier, Montpellier, France;}
\affil[3]{Service de Neurologie, Department of Clinical Neurosciences, Lausanne University Hospital (CHUV), Lausanne, Switzerland;}

\date{}

\begin{document}
\maketitle

\begin{abstract}
Accurate classification of neuronal cell types is essential for understanding brain organization, but multimodal neuron datasets are scarce and strongly imbalanced across subclasses. We present a benchmark of synthetic data augmentation methods for predicting electrophysiology-defined neuronal classes (e-types) in the Allen Cell Types mouse visual cortex dataset. Two supervised tasks were evaluated over the same 17 e-type labels: prediction from electrophysiology features alone (E\ensuremath{\to}e-type) and prediction from combined morphology plus electrophysiology features (M+E\ensuremath{\to}e-type). We established real-data baselines across multiple classifier families under a unified preprocessing pipeline, then augmented only the training sets using matched per-class grids with Synthetic Minority Over-sampling Technique (SMOTE) and deep generative models: Variational Autoencoders (VAE), Generative Adversarial Networks (GAN), masked autoregressive normalizing flows, and Denoising Diffusion Probabilistic Models (DDPM). Augmentation produced substantial generalization gains when applied in the native high-dimensional feature space, whereas introducing dimensionality reduction largely suppressed these benefits. SMOTE delivered the most robust and consistent improvements across tasks and augmentation levels. To assess biological realism, we introduced a fidelity framework combining feature-wise distribution comparisons, statistical concordance tests, and distance-based measures that compare synthetic-to-real variability against the natural variability between real classes. Most synthetic datasets stayed within biological diversity bounds, with deviations concentrated in the rarest subclasses. These results provide practical guidance on selecting and validating synthetic augmentation for neuronal subtype classification.
\end{abstract}

\section{Introduction}

The precise classification of neuronal cell types is fundamental for unraveling the structural and functional organization of the mammalian brain. High-throughput technologies such as patch-clamp electrophysiology and three-dimensional morphological reconstructions have facilitated profiling of neuronal properties, revealing a rich diversity of cell types in the cortex and beyond (Gouwens et al., 2019, 2020; Tasic et al., 2018; Vasques et al., 2016, 2023). Certain neuronal functions have been speculated to manifest exclusively at the circuit level (Luo, 2021), necessitating the utilization of connectivity properties when available. In other cases, neural activity alterations occur in a pathological or diseased state, and a decline in neural activity may be indicative of activity-dependent recruitment only of specific neuronal subpopulations, crucial for post-injury locomotion (Kathe et al., 2022). Structural properties of neurons, including dendritic and axonal morphologies, soma size, and spine density, can be integrated with electrophysiological features such as resting membrane potential, biophysical attributes, firing rate, as well as protein and mRNA composition, to comprehensively represent neuronal physiological and molecular properties (Cuntz et al., 2010; Fuzik et al., 2016; Nandi et al., 2022). A critical challenge remains in addressing the substantial cellular heterogeneity present within neural tissue, both in healthy and diseased states (Zeng, 2022). Functional categorization of distinct neuron types remains a pivotal task that demands multidimensional investigation at the level of individual neurons (Gouwens et al., 2019). Advancements in neuronal classification methodologies have been achieved through a multifaceted approach encompassing morphological, molecular, electrophysiological, transcriptomic (Scala et al., 2021; Yuste et al., 2020), and biophysical analyses, thereby enriching the systematic and reproducible understanding of biological structures and functions (Zhang et al., 2021). Comprehensive datasets remain scarce, particularly for rare or molecularly distinct subpopulations, and the intrinsic heterogeneity of neuronal phenotypes poses substantial challenges for developing robust and generalizable classification models (Scala et al., 2021; Yuste et al., 2020). Imbalanced datasets and insufficient representation of minority classes often limit the accuracy and reliability of machine learning approaches in neuroscience (Vasques et al., 2023). The advent of computational neuroscience methodologies has been paramount in quantifying the intricate interplay between neuronal structure and activity (Halavi et al., 2008; Torben-Nielsen et al., 2008). A thorough exploration of the morpho-physiological relationships in neurons necessitates extensive modeling to harmonize digitized neuron morphologies with physiological data, a process that demands meticulous manual effort. To accommodate the scarcity of available electrophysiological and morphological data from neurons, multiple generative approaches, pertaining especially to neuron morphologies, have been proposed (Ascoli \& Krichmar, 2000; Cuntz et al., 2010; Koene et al., 2009; Torben-Nielsen et al., 2008); these approaches use stringent a priori constraints and existing knowledge of neuronal morphology (Wanhainen \& Adamsson, 2021), which are potential sources of bias. Synthetic data generation has emerged as a promising solution to address these challenges. Classical oversampling techniques like the Synthetic Minority Over-sampling Technique (SMOTE) have been widely used to improve class balance in biological data (Chawla et al., 2002). More recently, deep generative models, including Generative Adversarial Networks (GANs), Variational Autoencoders (VAEs), Normalizing Flows, and Denoising Diffusion Probabilistic Models (DDPMs), have demonstrated remarkable success in learning complex, high-dimensional distributions, and generating synthetic data that closely mimic real biological variability (Hartmann et al., 2018; Ho et al., 2020; Kingma \& Welling, 2013; Molano-Mazon et al., 2018; Papamakarios et al., 2019). In neuroscience, these models have enabled simulation of electrophysiological recordings (Hartmann et al., 2018), spike trains (Molano-Mazon et al., 2018), and even high-dimensional morphological and transcriptomic features (Gouwens et al., 2020; Marouf et al., 2020), contributing to improved classifier performance.

Despite these advances, the field still lacks standardized and biologically grounded criteria for validating the fidelity of synthetic neuronal data, particularly for morpho-electrophysiological profiles. Most published studies evaluating generative models in neuroscience have focused on visual inspection, distributional similarity of selected features, or improvements in downstream classifier performance as primary measures of synthetic data quality (Gouwens et al., 2020; Hartmann et al., 2018; Molano-Mazon et al., 2018). However, such metrics often overlook the fundamental requirement that synthetic neurons should reproduce not only the central tendencies, but also the natural phenotypic diversity and boundaries characterizing real neuronal populations. Only a handful of recent studies, mainly in transcriptomics and spike train modeling, have begun to emphasize the need for benchmarks that directly compare synthetic-to-real variability against the natural variability between real classes or cell types (Marouf et al., 2020). For morpho-electrophysiological neurons in particular, this gap is even more pronounced: while classification frameworks have been established using combined morphological and electrophysiological data (Gouwens et al., 2019; Scala et al., 2021), comprehensive guidelines or best practices for evaluating synthetic data fidelity in this domain are essentially absent from the literature. Consequently, there remains a need for biologically meaningful benchmarking strategies that can distinguish between synthetic datasets that merely mimic superficial characteristics and those that genuinely recapitulate the full spectrum of natural neuronal diversity. Synthetic neurons can play a pivotal role in the preliminary phases of drug discovery and electroceuticals, enabling initial in silico evaluations on artificial models prior to in vivo implementations and aiding the exploration of neurological and neurodegenerative conditions and potential remedial interventions. In this study, we carried out a benchmark of synthetic data augmentation for neuronal subtype prediction using the Allen Cell Types mouse dataset, in which each neuron is annotated by intrinsic electrophysiology alone or by joint morphology plus electrophysiology (Gouwens et al., 2019, 2020). We formulated two supervised tasks over the same e-type label space, i.e., electrophysiological cell types defined in the Allen Cell Types taxonomy from intrinsic firing phenotypes (such as regular-spiking adapting excitatory vs fast-spiking inhibitory patterns) (Gouwens et al., 2019, 2020). Specifically, E\ensuremath{\to}e-type predicts these e-types from electrophysiological features only (intrinsic firing and membrane properties), whereas M+E\ensuremath{\to}e-type predicts the same e-types from the joint morpho-electrophysiological feature set (3D reconstructions plus intrinsic physiology). For each task, we established real-data baselines across multiple classifier families under a unified preprocessing pipeline. For augmentation, we generated class-conditional synthetic neurons with both a classical oversampling method (SMOTE) and deep generative approaches (VAE, GAN, masked autoregressive flow, and DDPM), following a controlled per-class augmentation schedule (increasing synthetic counts over several low-to-high regimes) applied only to the training set. We further evaluated augmentation in two settings: directly in the native feature space and after dimensionality reduction, mirroring common practice in morpho-electrophysiological classification. Beyond reporting downstream gains, we introduced a biologically grounded validation framework to quantify synthetic fidelity. This framework combines qualitative real-vs-synthetic feature-distribution comparisons, feature-wise statistical concordance tests, and distance-based measures that compare synthetic-to-real variability against the natural variability observed between real neuronal classes. Together, these analyses allowed us to distinguish augmentation methods that merely improve classifiers from those that genuinely preserve the phenotypic structure and diversity of cortical e-types.

\section{Results}

Before introducing synthetic data, we established baselines to quantify how well different feature sets discriminate neuronal subclasses (Figure 1). In all analyses, the labels are 17 electrophysiology-defined neuron types (e-types) from the Allen Cell Types dataset: 4 excitatory classes (Exc\_1--Exc\_4) and 13 inhibitory classes (Inh\_1--Inh\_13) obtained by unsupervised clustering and aligned with transcriptomic subclasses (Gouwens et al., 2019). Each short label is a cluster ID with a characteristic firing profile---for example, Exc\_3 corresponds to a regular-spiking adapting excitatory neuron (“RS adapt.”) and Inh\_10 to a fast-spiking sustained inhibitory interneuron (“FS sust. 1”)---and our classifiers always predict these underlying e-type IDs (Figure 2).

We evaluated two supervised tasks that share this same label space but differ in their inputs: (i) E\ensuremath{\to}e-type, which uses only intrinsic electrophysiological features, and (ii) M+E\ensuremath{\to}e-type, which uses combined morphology + electrophysiology features for the subset of neurons with both modalities available. Thus, morphology is treated as additional input information rather than a separate labeling axis, and the M+E\ensuremath{\to}e-type dataset simply denotes the 451 neurons with both modalities, all retaining their original 17 e-type labels. Overall, the E\ensuremath{\to}e-type dataset comprises 1,857 neurons (885 excitatory, 972 inhibitory), and the M+E\ensuremath{\to}e-type dataset comprises 451 neurons (232 spiny/putative excitatory, 219 aspiny or sparsely spiny/putative inhibitory). Supplementary Table 1 details the per-class counts.

For each task, we applied a standardized preprocessing pipeline (feature curation, stimulus-relative timing, and feature-family-specific scaling), optional dimensionality reduction, and a suite of supervised classifiers. Figure 1 summarizes the end-to-end workflow from preprocessing and rescaling through model training and evaluation. These baselines establish the reference performance against which we later compare the added value of synthetic data augmentation.

\begin{figure}[t]
                \centering
                \includegraphics[width=\linewidth]{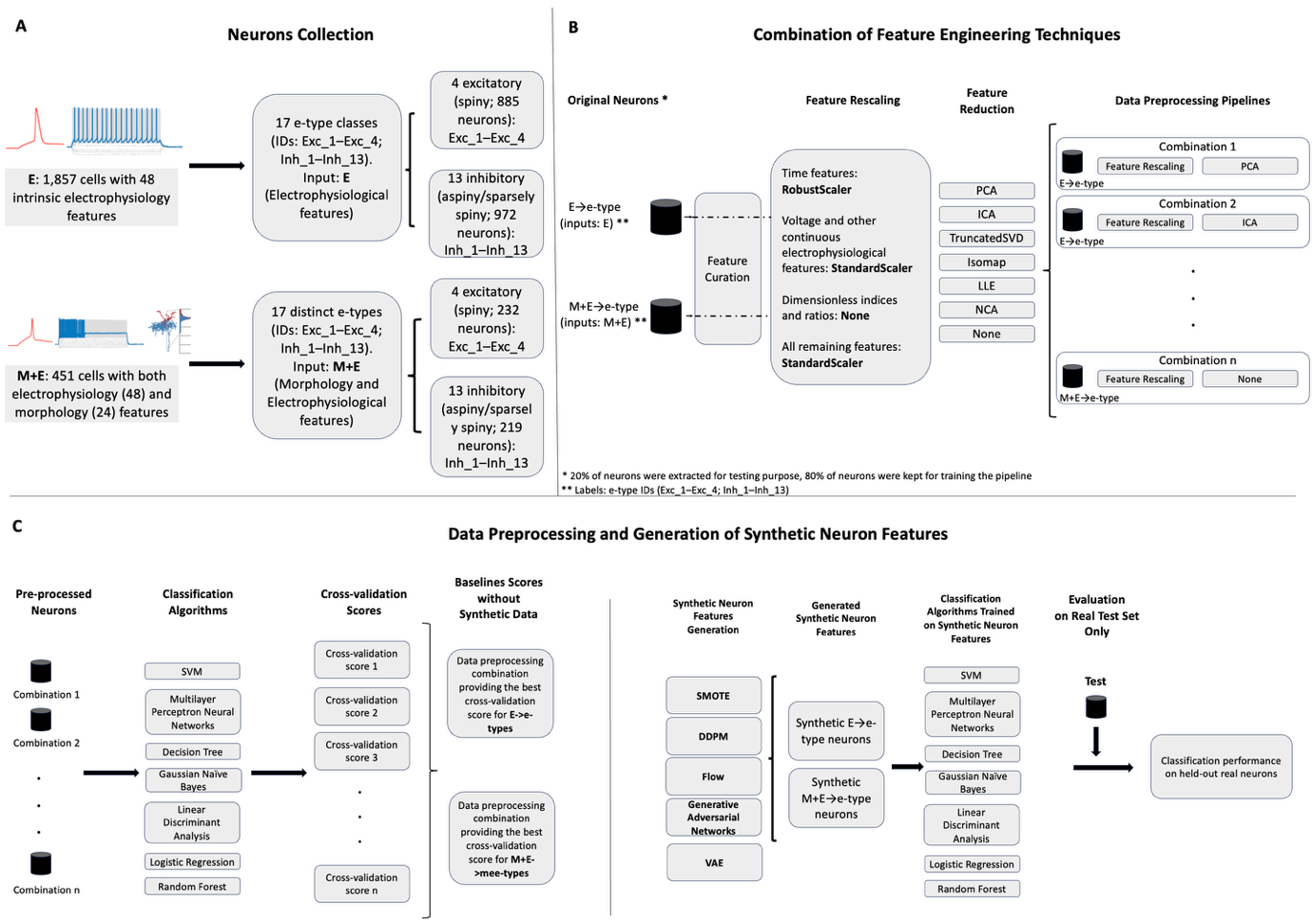}
                \caption{Overview of datasets, preprocessing, and synthetic feature generation for e-type classification. (A) Neuron collection and label space. The E dataset comprises 1,857 neurons with 48 intrinsic electrophysiological features; the M+E dataset comprises 451 neurons with both electrophysiological (n=48) and morphology-derived (n=24) features. In all analyses, the labels are the 17 electrophysiology-defined neuron classes (e-types) from the Allen Cell Types database (4 excitatory classes, Exc\_1--Exc\_4; 13 inhibitory classes, Inh\_1--Inh\_13), as in Gouwens et al. (Gouwens et al., 2019). Excitatory vs inhibitory counts are indicated for each dataset. Morphology is used only as additional input in the M+E\ensuremath{\to}e-type task; the label space is identical in both settings.
(B) Preprocessing and feature engineering. For each dataset, we applied explicit feature curation to remove strictly technical variables, followed by a fixed ColumnTransformer that performs family-specific scaling (RobustScaler for timing features, StandardScaler for voltage and other continuous variables, no rescaling for ratios/indices, and StandardScaler for morphology metrics), and an optional dimensionality reduction step (PCA, ICA, TruncatedSVD, Isomap, LLE, NCA, or no reduction). Combinations of scaler + reducer define the preprocessing pipelines evaluated downstream.
(C) Classification baselines and synthetic feature generation. Each preprocessing pipeline was combined with multiple supervised classifiers (SVM, multilayer perceptron, decision tree, random forest, Gaussian Naïve Bayes, linear discriminant analysis, logistic regression) and evaluated by stratified cross-validation to select robust baselines for E\ensuremath{\to}e-type and M+E\ensuremath{\to}e-type. Preprocessing + classifier configurations were then reused to train models on augmented training sets in which real neurons were supplemented with synthetic samples generated by SMOTE, DDPM, normalizing flows, GANs, or VAEs. In all cases, final performance was assessed on the same held-out real test set, allowing us to quantify the impact of synthetic data on the recovery of e-type classes.}
                \label{fig:fig1}
                \end{figure}

\subsection{Comprehensive Evaluation of Classification Pipelines: Establishing the Baseline Landscape}

To establish a foundation for subsequent analyses, we systematically evaluated a grid of supervised learning pipelines for neuron type classification based on electrophysiological features alone (E\ensuremath{\to}e-type) and on combined morpho-electrophysiological features (M+E\ensuremath{\to}e-type). Each pipeline was defined as a unique combination of feature rescaling, optional dimensionality reduction, and classifier, and was tuned by stratified cross-validation on the training split. The held-out test set was used once, only for final evaluation of the selected baselines (and later for evaluation with synthetic data). Across both tasks, the benchmark comprised 7,040 evaluated pipelines in total: 3,520 for E\ensuremath{\to}e-type and 3,520 for M+E\ensuremath{\to}e-type, each spanning conditions with and without dimensionality reduction. For every configuration we recorded multiple performance metrics---including accuracy, macro-F1, macro-precision, and macro-recall---to account for class imbalance and heterogeneous error costs. All cross-validated scores are reported in the Supplementary Material: Baseline.

To quantify generalization and avoid severely overfitting models, we examined the train--test accuracy gap for each pipeline and retained only those with an absolute gap \ensuremath{\leq} 0.07 for visualization in Figure 3A--D. This conservative filter removed the most overconfident configurations while preserving a large and diverse set of well-behaved models. For E\ensuremath{\to}e-type, 1,946 of 3,520 pipelines (55.3\%) met this criterion, including 14/320 (4.4\%) without dimensionality reduction and 1,932/3,200 (60.4\%) with a reducer. For M+E\ensuremath{\to}e-type, 1,855 of 3,520 pipelines (52.7\%) were retained, including 11/320 (3.4\%) without reduction and 1,844/3,200 (57.6\%) with a reducer. Figure 3A shows the resulting performance landscape: for each dataset and condition, points are positioned by mean test accuracy (x-axis) and train--test gap (y-axis), revealing a continuum from underfitting to mild overfitting.

\begin{figure}[t]
                \centering
                \includegraphics[width=\linewidth]{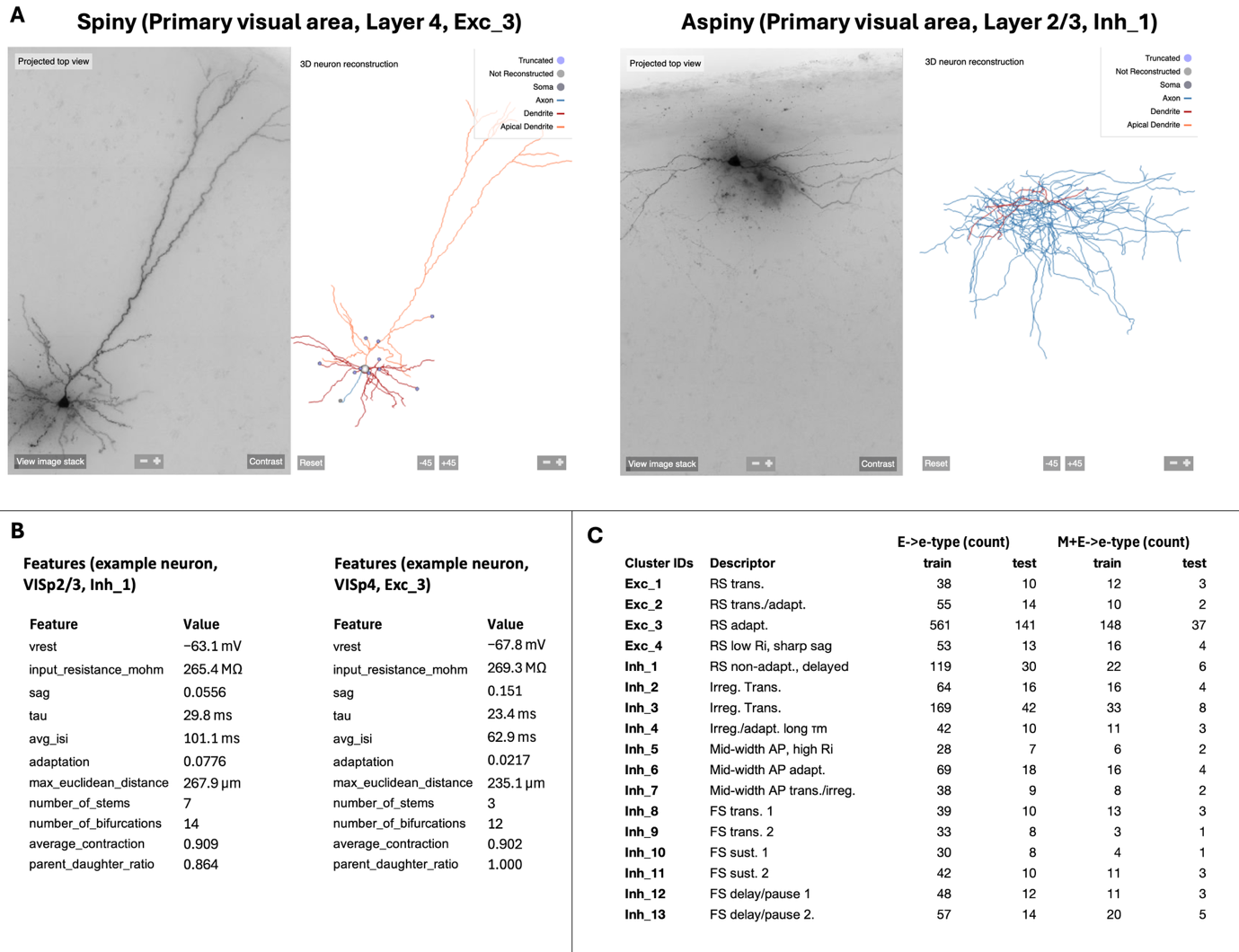}
                \caption{Example neurons, features, and class structure used for supervised classification.
(A) Representative excitatory and inhibitory neurons from the Allen Cell Types Database with both electrophysiological and morphological recordings. Left: spiny pyramidal cell from primary visual cortex layer 4 (VISp4, e-type Exc\_3), shown as a projected image stack (left panel) and 3D reconstruction (right panel; soma in gray, dendrites in red, axon in blue). Right: aspiny interneuron from VISp layer 2/3 (VISp2/3, e-type Inh\_1) shown with the same visualizations. These examples illustrate the contrasting dendritic and axonal arborizations of excitatory versus inhibitory neurons in mouse visual cortex. (B) Subset of electrophysiological and morphological features used as model inputs for the two example cells in panel A. For each neuron, we report key intrinsic properties derived from standardized current-clamp protocols (e.g. resting membrane potential vrest, input resistance, sag ratio, membrane time constant \ensuremath{\tau}, mean inter-spike interval avg\_isi, and spike-frequency adaptation index), together with summary morphology metrics (e.g. maximal Euclidean path length from soma, number of primary stems and bifurcations, average contraction, and parent--daughter diameter ratio). These features are directly drawn from the Allen Cell Types feature matrix and form part of the 48 electrophysiological and 24 morphological variables used throughout the study.
(C) Electrophysiological class labels (e-types) used as prediction targets. The 17 classes comprise 4 excitatory clusters (Exc\_1--Exc\_4) and 13 inhibitory clusters (Inh\_1--Inh\_13), each summarized by a plain-English descriptor of its firing pattern and intrinsic properties (e.g. regular spiking transient/adapting, irregular spiking, fast-spiking sustained, FS delay/pause). The right columns report the number of neurons per class in the training and test sets for the electrophysiology-only (E\ensuremath{\to}e-type) and morpho-electrophysiology (M+E\ensuremath{\to}e-type) datasets.}
                \label{fig:fig2}
                \end{figure}

\begin{figure}[p]
  \centering
  \includegraphics[width=\linewidth,height=0.78\textheight,keepaspectratio]{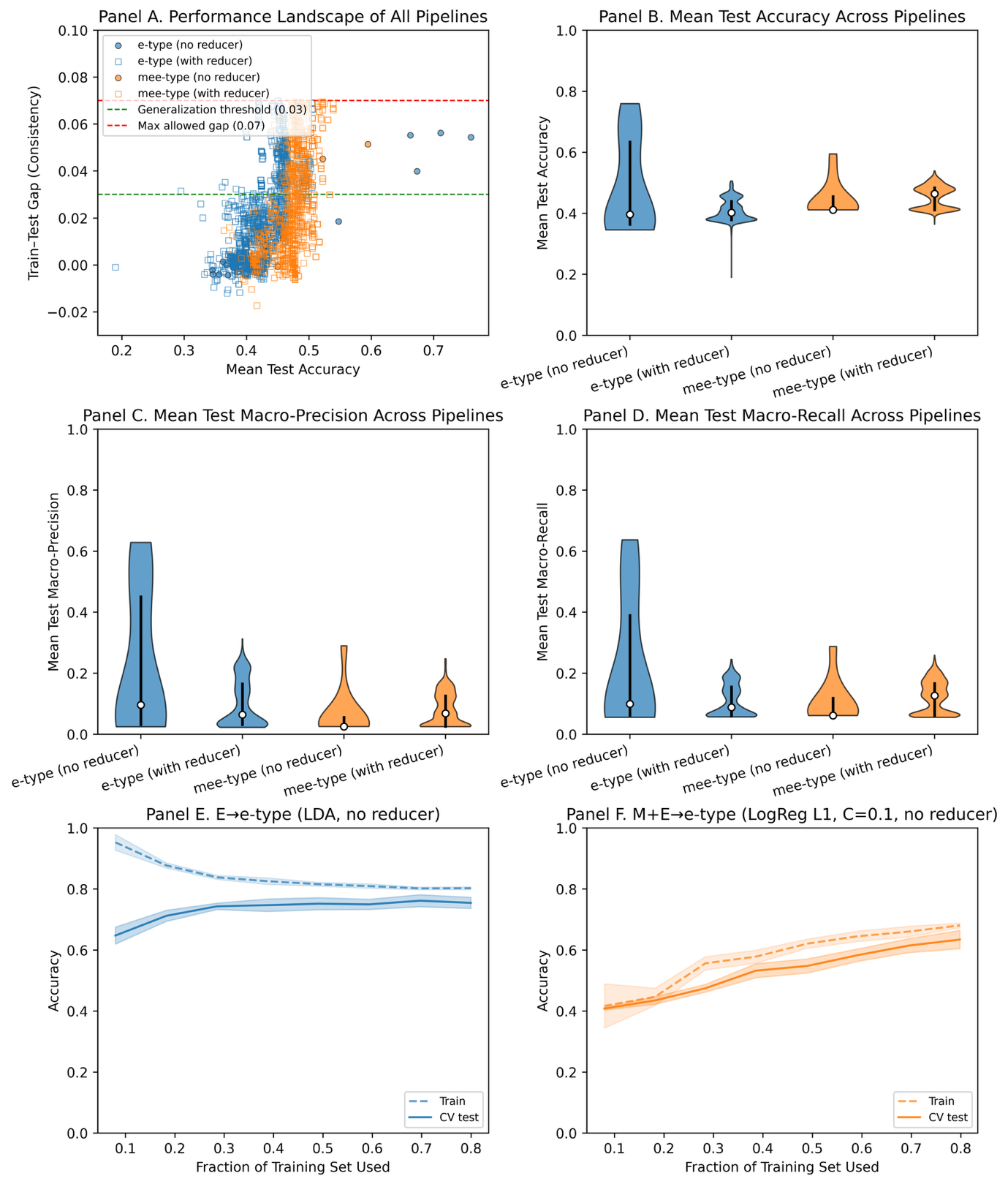}
  \caption{Baseline performance landscape and learning curves for neuronal subtype classification.}
  \label{fig:fig3}
\end{figure}

Within this filtered set, the best electrophysiology-only pipelines without dimensionality reduction achieved the highest test accuracies. The top E\ensuremath{\to}e-type model, a linear discriminant analysis (LDA) classifier without reducer, reached a mean cross-validated test accuracy of 0.76 (\ensuremath{\pm}0.02), with macro-F1 \ensuremath{\approx} 0.62, macro-precision \ensuremath{\approx} 0.63, and macro-recall \ensuremath{\approx} 0.64, and a moderate train--test gap of \ensuremath{\approx} 0.05. In contrast, the best M+E\ensuremath{\to}e-type model without reducer, an L1-regularized logistic regression (C = 0.1), reached a mean test accuracy of 0.59 (\ensuremath{\pm}0.06), with macro-F1 \ensuremath{\approx} 0.28, macro-precision \ensuremath{\approx} 0.29, and macro-recall \ensuremath{\approx} 0.29, and a similar gap (\textasciitilde{}0.05). Thus, even when using balanced metrics, electrophysiology-only models not only reach higher absolute performance but also obtain more favorable macro-scores than their morpho-electrophysiological counterparts. Figure 3B--D summarizes the distributions of mean test accuracy, macro-precision, and macro-recall across all pipelines with |gap| \ensuremath{\leq} 0.07: for E\ensuremath{\to}e-type without dimensionality reduction, the distributions are broad and shifted toward higher values, whereas all conditions involving M+E\ensuremath{\to}e-type features show lower medians and left-shifted, more compact distributions. These patterns confirm that the advantage of E\ensuremath{\to}e-type over M+E\ensuremath{\to}e-type is not an artifact of relying solely on accuracy.

For downstream synthetic data experiments, we further tightened our baseline selection to focus on highly stable and conservative classifiers. Starting from the same grid, we restricted attention to pipelines with an absolute train--test accuracy gap \ensuremath{\leq} 0.03 and then selected, separately for each dataset (E and M+E) and condition (with or without reducer), the configuration with the highest mean test accuracy under this stricter constraint. This procedure yielded four baselines (E\ensuremath{\to}e-type with and without reduction; M+E\ensuremath{\to}e-type with and without reduction), which are summarized in Table 1.

The selected baselines (Table 1) confirm that electrophysiology alone is already informative for e-type prediction, but that performance remains moderate and strongly class-imbalanced. For E\ensuremath{\to}e-type, the best SVM with RBF kernel (no reducer) reaches a mean test accuracy of 0.55 \ensuremath{\pm} 0.09, but the macro-F1 score is only 0.26, indicating that rare subclasses are still poorly captured. Introducing NCA before an MLP slightly reduces accuracy (0.50 \ensuremath{\pm} 0.03) and macro-F1 (0.19), suggesting that dimensionality reduction does not help the baseline electrophysiological models. For M+E\ensuremath{\to}e-type, the SVM without reducer performs worst in terms of macro-F1 (0.07), while the LLE + random forest pipeline improves both accuracy (0.53 \ensuremath{\pm} 0.02) and macro-F1 (0.20), yet still lags behind the best E\ensuremath{\to}e-type baseline. Overall, these baselines provide conservative but clearly improvable starting points, motivating the use of synthetic data generation to enhance performance and balance across neuronal subclasses.

\begin{table}[t]
\centering
\caption{Baseline classification pipelines for electrophysiological and morpho-electrophysiological neuron type prediction. Baseline models were selected separately for the E\ensuremath{\to}e-type (electrophysiology only) and M+E\ensuremath{\to}e-type (morphology + electrophysiology) tasks, with and without dimensionality reduction. For each condition, we retained the pipeline with the highest mean cross-validated test accuracy among those with a train--test accuracy gap \ensuremath{\leq} 0.03. The table reports mean test accuracy (\ensuremath{\pm} standard deviation across folds), test macro-averaged F1 score, precision and recall, mean train accuracy (\ensuremath{\pm} standard deviation), the train--test accuracy gap, and the main classifier hyperparameters, together with the chosen reducer and classifier type. These baselines are used as reference models for evaluating the impact of synthetic data augmentation in subsequent analyses.}

\label{tab:tab1}
\small
\setlength{\tabcolsep}{4pt}
\renewcommand{\arraystretch}{1.15}
\resizebox{\linewidth}{!}{%
\begin{tabular}{lccccccc p{6.8cm} cc}
\toprule
Dataset & Mean Test Accuracy & Mean Test F1 & Mean Test Precision & Mean Test Recall & Mean Train Accuracy & Gap Train--Test & Parameters & Reducer & Classifier \\
\midrule
E\ensuremath{\to}e-type & $0.55\pm0.09$ & 0.26 & 0.34 & 0.26 & $0.57\pm0.09$ & 0.02 &
\texttt{\{classifier\_C: 10, classifier\_gamma: 'scale'\}} & none & svm\_rbf \\
E\ensuremath{\to}e-type & $0.50\pm0.03$ & 0.19 & 0.22 & 0.20 & $0.52\pm0.03$ & 0.03 &
\texttt{\{classifier\_activation: 'tanh', classifier\_alpha: 0.0001, classifier\_hidden\_layer\_sizes: (128,), classifier\_learning\_rate\_init: 0.0003\}} & nca & mlp \\
M+E\ensuremath{\to}e-type & $0.46\pm0.02$ & 0.07 & 0.05 & 0.12 & $0.47\pm0.003$ & 0.01 &
\texttt{\{classifier\_C: 1, classifier\_gamma: 'scale'\}} & none & svm\_rbf \\
M+E\ensuremath{\to}e-type & $0.53\pm0.02$ & 0.20 & 0.21 & 0.23 & $0.56\pm0.01$ & 0.03 &
\texttt{\{classifier\_max\_depth: None, classifier\_min\_samples\_leaf: 5, classifier\_min\_samples\_split: 20, classifier\_n\_estimators: 50, reducer\_n\_neighbors: 10\}} & lle & rf \\
\bottomrule
\end{tabular}%
}
\end{table}

\subsection{Synthetic Data Generation for Mouse Visual Cortex Neurons Based on Electrophysiological and Morphological Features with and without reducer}

To quantify how synthetic data influences generalization to entirely real held-out neurons, we evaluated all augmentation strategies using Holdout Accuracy, ensuring that improvements reflect genuine external validity rather than cross-validation artifacts (Supplementary Material: Synthetic). Figure 4 summarizes the performance landscape across the unified augmentation grid (N = 0, 100, 500, 1000, 5000, 10000 synthetic neurons per class), jointly examining electrophysiology-only classification (E\ensuremath{\to}e-type), multimodal morphology+electrophysiology classification (M+E\ensuremath{\to}e-type), and the role of dimensionality reduction.

\begin{figure}[t]
                \centering
                \includegraphics[width=\linewidth]{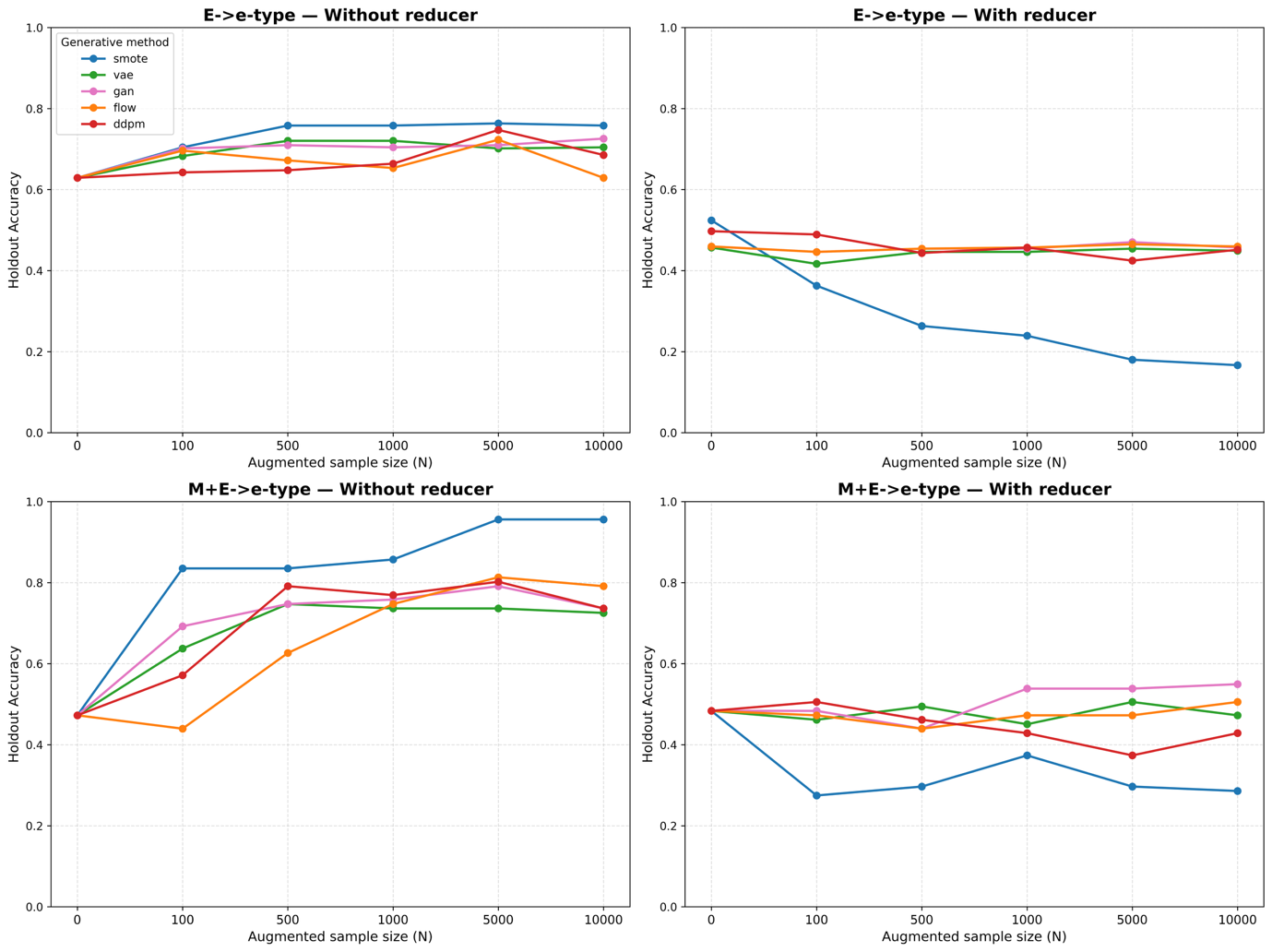}
                \caption{Holdout Accuracy for neuronal subtype classification with synthetic data augmentation.
Performance is shown for electrophysiology-only neurons (E\ensuremath{\to}e-type, top row) and multimodal morphology+electrophysiology neurons (M+E\ensuremath{\to}e-type, bottom row), across increasing amounts of synthetic data (N = 0, 100, 500, 1000, 5000, 10000). Left panels display results obtained in the native feature space without dimensionality reduction; right panels show results when PCA, ICA, SVD, or NCA were applied before classification. Each curve corresponds to one augmentation method (SMOTE, VAE, GAN, Normalizing Flow, DDPM).}
                \label{fig:fig4}
                \end{figure}

Synthetic augmentation robustly improves generalization only in the native feature space, that is, when no dimensionality reduction is applied. Across both E\ensuremath{\to}e-type and M+E\ensuremath{\to}e-type tasks, augmentation in the full electrophysiological feature space markedly improves hold-out performance (Figure 4, left panels), and this improvement is substantial when compared directly to the real-data baselines. For E\ensuremath{\to}e-type, the baseline hold-out accuracy obtained from real data alone (baseline) is approximately 0.55, and synthetic augmentation consistently surpasses this level once N increases beyond 100. SMOTE provides the most reliable gains across all augmentation regimes. For E\ensuremath{\to}e-type, SMOTE exhibits a monotonic improvement with increasing N, stabilizing between 0.72 and 0.76 once N \ensuremath{\geq} 1000. For M+E\ensuremath{\to}e-type, the gains are even more pronounced, increasing Holdout Accuracy from baseline values around 0.46 (and up to 0.53 when using LLE+RF) to nearly 0.90 for N = 5000--10000. This strong improvement is consistent with the observation about the low sample-to-feature ratio without augmentation, suggesting that synthetic augmentation effectively regularizes the multimodal manifold, densifying the feature space and stabilizing class boundaries that are poorly supported by real data alone.

Deep generative models also yield moderate but consistent improvements. GANs, Normalizing Flows, and DDPM models produce smoother performance curves but rarely reach the peak values achieved with SMOTE. Their benefits are clearest in classes with broad electrophysiological variability, such as regular-spiking adapting excitatory neurons, and in morphological subclasses where dendritic and axonal measures contribute nonlinear discriminative cues. Importantly, these models perform consistently above the N = 0 baseline, demonstrating that deep synthetic neurons remain sufficiently biologically plausible to support generalization. When compared to the real-data baselines, even the more modest improvements from GAN, Flow, VAE, or DDPM still represent an absolute increase of 0.10 to 0.20 in Holdout Accuracy, confirming that augmentation meaningfully enhances classifier stability.

Augmentation benefits saturate at N \ensuremath{\approx} 1000--5000. For both modalities, the transition from N = 0 to N = 100 yields modest gains corresponding to a data-diversity regime, followed by substantial improvements between N = 100 and N = 1000, marking the manifold-stabilization regime. Between N = 5000 and N = 10000, gains plateau or fluctuate mildly due to class-dependent imbalance. This plateau reinforces that augmentation volume is not the dominant driver of performance once a critical density is reached, in line with recent theoretical work on generative augmentation in high-dimensional manifolds.

In contrast with the clear improvements observed in the native feature space, applying PCA, ICA, TruncatedSVD, or NCA before training substantially alters the effect of augmentation (Figure 4, right panels). For the E\ensuremath{\to}e-type task, dimensionality reduction consistently degrades SMOTE performance as N increases, with Holdout Accuracy dropping from \textasciitilde{}0.48 at N = 0 to \textasciitilde{}0.16 at N = 10000. Deep generative models (GAN, Flow, VAE, DDPM) do not collapse but instead remain relatively flat, fluctuating between 0.40 and 0.50 across all augmentation levels. Although these values do not show catastrophic degradation, none of the methods yield improvements over the unreduced pipelines, and several configurations fall below their corresponding real-data baselines.

For the M+E\ensuremath{\to}e-type task, the patterns are more heterogeneous. Most augmentation methods remain in a narrow band between 0.40 and 0.55 and do not exhibit clear benefits from increased synthetic data. Overall, dimensionality reduction reduces the consistency and reliability of augmentation effects, and in most cases suppresses the performance gains observed in the native feature space.

There is a clear rationale for why augmentation fails after reduction. The electrophysiological features that define E->e-types---such as sag ratio, spike width, latency to first spike, and firing rate adaptation---are not globally correlated; their discriminative value is distributed in a nonlinear, high-dimensional manner. Projection to a linear subspace mixes fast-spiking and regular-spiking phenotypes, compresses latency-based descriptors, suppresses morphology-specific variation in the M+E\ensuremath{\to}e-type task, and removes subtle cross-feature interactions necessary for class resolution. This explains why augmentation in the native feature space is beneficial, while augmentation in a reduced space becomes ineffective or harmful, and why reduced-space performance often falls below the real-data baseline.

Comparing E\ensuremath{\to}e-type and M+E\ensuremath{\to}e-type conditions shows that the native electrophysiological space stabilizes earlier. The SMOTE curve for E\ensuremath{\to}e-type saturates at N \ensuremath{\approx} 1000--2000, reflecting the relatively higher density of electrophysiology-only data and the strong low-dimensional structure governing spike timing, excitability, and adaptation. By contrast, multimodal M+E\ensuremath{\to}e-type benefits more from synthetic augmentation: it starts from a lower unaugmented baseline, exhibits stronger gains with synthetic data, shows a delayed saturation point around N \ensuremath{\approx} 5000--10000, and demonstrates larger improvements for GAN, Flow, and DDPM, consistent with the richer structure in the morphology subspace. These observations confirm that the multimodal manifold is under-sampled and benefits disproportionately from synthetic densification.

Altogether, the trends in Figure 4 point to several methodological implications. Synthetic augmentation is effective only when applied before dimensionality reduction, emphasizing that synthetic neurons must be generated and evaluated in the full native feature space where the discriminative structure resides. SMOTE remains the most reliable augmentation strategy across both modalities and all augmentation sizes, yielding stable, monotonic, and biologically consistent improvements that far surpass the real-data baselines. Deep generative models also improve generalization, though more modestly, and their optimal sample size varies by task.

\subsection{Qualitative and Quantitative Assessment of Augmentation Strategies}

To evaluate the fidelity and biological plausibility of the synthetic neurons generated in this study, we developed a validation framework combining both qualitative and quantitative criteria. This framework assesses how well synthetic samples reproduce the statistical structure, class-specific signatures, and natural variability of electrophysiological and morpho-electrophysiological neuron types.

The evaluation proceeds along four complementary axes. Feature-wise distributional similarity is quantified using class-conditional Kolmogorov--Smirnov (KS) tests, allowing us to assess how closely synthetic neurons match the empirical distributions of real features for each neuronal class. The class-wise fidelity of mean phenotypic profiles is assessed through mean absolute error (MAE) and Euclidean distances between synthetic and real class means, providing insight into how accurately each augmentation strategy reproduces characteristic electrophysiological and morphological signatures. Generalization fidelity is evaluated by comparing synthetic-to-real MAE with the natural phenotypic variability observed among real neuron classes. This establishes whether synthetic diversity remains within biologically plausible boundaries. Finaly, feature-level divergence mapping, enabled by heatmaps and merged real-versus-synthetic distributional plots, identifies the specific features and neuronal subclasses that are best captured by synthetic data, as well as those exhibiting systematic mismatches.

\subsubsection{Qualitative Comparison of Real and Synthetic Neuron Distributions}

To qualitatively assess how well synthetic samples reproduce the empirical structure of real neuronal populations, we examined merged kernel-density distribution plots for all electrophysiological and morpho-electrophysiological features (Figure 5). These visualizations compare, feature by feature, the density profiles of real training neurons, real held-out test neurons, and synthetic neurons generated with SMOTE at 5,000 samples per class for both E\ensuremath{\to}e-type and M+E\ensuremath{\to}e-type datasets.

Across most features, synthetic distributions closely followed the empirical shape and scale of real neurons, indicating that SMOTE preserves key aspects of neuronal variability when applied at large sample sizes. Several features, particularly those related to subthreshold membrane properties (e.g., sag, input resistance) and action-potential timing (e.g., peak\_t, trough\_t), showed near-perfect overlap between synthetic and real densities for the majority of classes. This suggests that interpolation-based oversampling is well suited to capturing smooth, unimodal relationships commonly observed in these electrophysiological dimensions.

For features exhibiting broader or multimodal distributions in biological neurons (e.g., adaptation, latency, upstroke--downstroke ratios), synthetic samples reproduced the global structure but occasionally displayed sharper peaks or reduced tail variability, consistent with SMOTE’s tendency to concentrate synthetic points along locally linear manifolds. These effects remained limited for E\ensuremath{\to}e-type but became more pronounced for M+E\ensuremath{\to}e-type, where several morphological features show intrinsically higher variance and heavier tails across neuron classes.

Morphological metrics (e.g., total length, number of bifurcations, overall depth) exhibited high qualitative fidelity overall, with synthetic curves aligning with both the central mass and tail behavior of the real distributions. A small subset of high-variance features presented slight misalignments between synthetic and real densities, consistent with quantitative results reported below.

Importantly, for both datasets, real test distributions consistently overlapped with synthetic data at least as well as with real training data. This observation indicates that synthetic points remain well contained within the natural biological variability of each neuron type and do not introduce spurious modes or unrealistic feature ranges.

\begin{figure}[p]
  \centering
  \includegraphics[width=\linewidth,height=0.78\textheight,keepaspectratio]{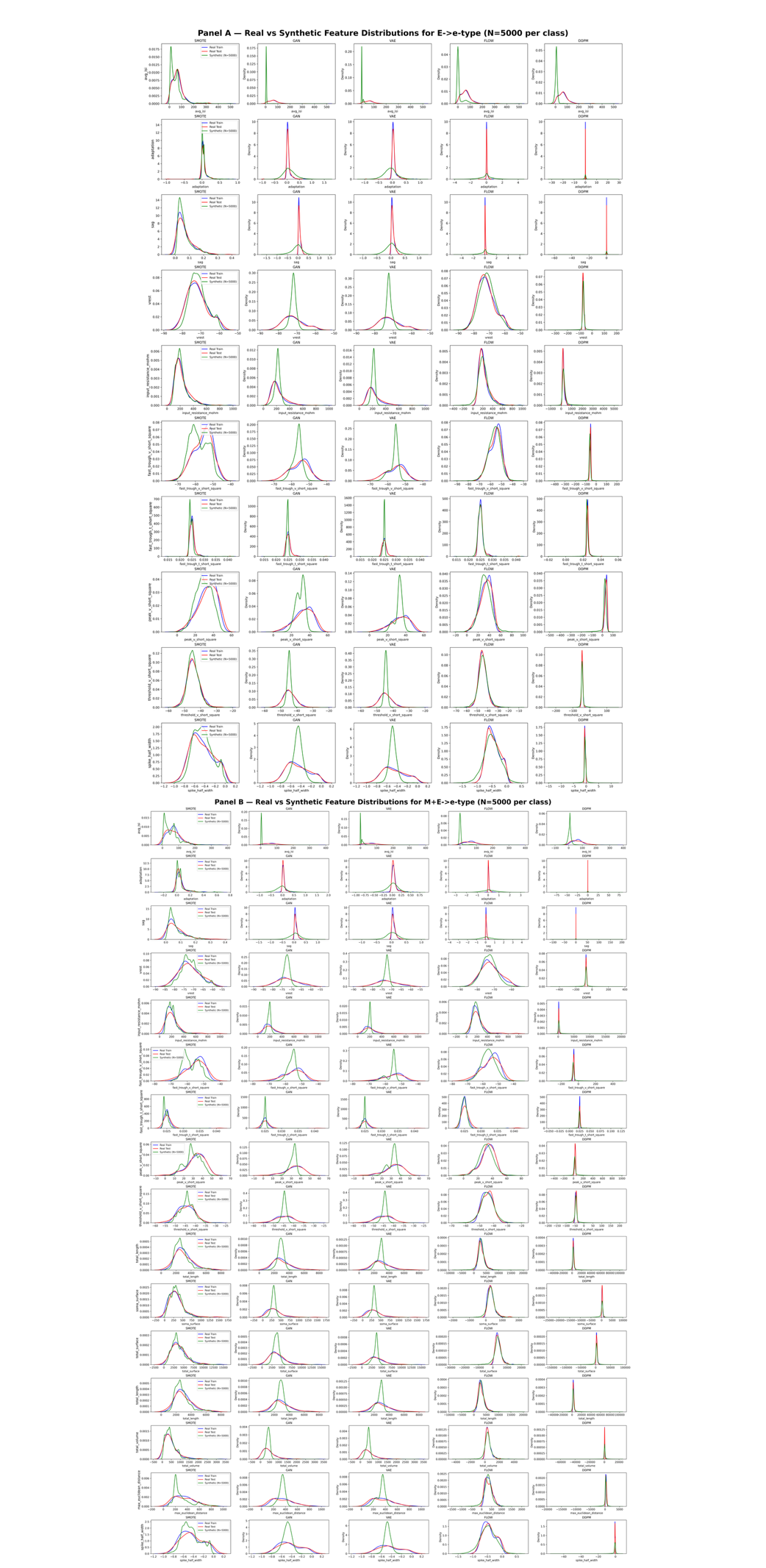} 
  \caption{Merged feature-wise kernel density estimates comparing real training data (blue), real test data (red), and synthetic neurons generated with SMOTE at 5,000 samples per class (green) for both electrophysiological (E\ensuremath{\to}e-type; Panel A) and morpho-electrophysiological datasets (M+E\ensuremath{\to}e-type; Panel B). Each subplot represents the empirical distribution of a single electrophysiological or morphological feature.}
  \label{fig:fig5}
\end{figure}

\subsubsection{Feature-wise Distributional Similarity (Kolmogorov--Smirnov Analysis)}

To quantify how closely synthetic neurons reproduce the empirical variability of real electrophysiological and morpho-electrophysiological features, we conducted class-conditional Kolmogorov--Smirnov (KS) tests comparing the distribution of each feature in the real data to its synthetic counterpart. This analysis was performed independently for the E\ensuremath{\to}e-type and M+E\ensuremath{\to}e-type datasets and for both augmentation levels (N = 5000 and N = 10000 synthetic samples per class). Across all conditions, the synthetic distributions closely matched those of real neurons, with only a small number of class--feature combinations exhibiting moderate divergence.

For the electrophysiology-only dataset (E\ensuremath{\to}e-type), KS statistics remained consistently low. Under the SMOTE 5000 condition, the average KS statistic was approximately 0.10, well within the range typically interpreted as high distributional similarity, with most features showing nearly identical cumulative distributions between real and synthetic neurons. For instance, features capturing spike dynamics such as fast\_trough\_v\_short\_square or peak\_t\_short\_square displayed very small divergences across most cell types, including well-represented classes like Exc\_3, Inh\_6, and Inh\_13. Even more variable features, such as adaptation index or avg\_isi, exhibited KS statistics indicative of strong alignment, particularly in classes with abundant real samples. Only a few inhibitory subclasses, notably Inh\_9 and Inh\_11, showed isolated instances of elevated KS statistics (e.g., 0.20--0.24 on features associated with firing latency or threshold), reflecting their intrinsically higher variability rather than systematic distortions in the synthetic data. Increasing the synthetic quantity to SMOTE 10000 did not degrade fidelity: the full distribution of KS values, including mean, median, and upper tail, remained essentially unchanged.

The morpho-electrophysiological dataset (M+E\ensuremath{\to}e-type) produced a similar overall pattern, despite the increased dimensionality introduced by morphology. Here, KS statistics averaged around 0.18 for SMOTE 5000 synthetic neurons. Many electrophysiological features, including vrest, input\_resistance\_mohm, and spike timing metrics, showed high correspondence across nearly all classes. Importantly, several morphological descriptors, such as soma\_surface, number\_bifurcations, and total\_length, also exhibited strong distributional similarity, particularly in inhibitory subclasses with richer morphological sampling (e.g., Inh\_1, Inh\_5, Inh\_6, Inh\_13). Divergences were again localized rather than systematic: for example, Inh\_8 and Exc\_2 showed somewhat elevated KS values on a handful of high-variance morphology features, such as average\_diameter or overall\_height, reflecting class-specific morphological heterogeneity rather than deficiencies in the generative model. As in the electrophysiology-only condition, doubling the synthetic sample size to SMOTE 10000 did not introduce additional discrepancies, and the distribution of KS statistics remained effectively unchanged.

Taken together, the KS analyses demonstrate that SMOTE preserves the univariate feature distributions of neuronal classes with a high degree of fidelity. The synthetic neurons emulate both electrophysiological and morphological variability, even within features that are known to be biologically noisy or class-specific. Deviations---when present---are concentrated in a small number of inhibitory subclasses (e.g., Inh\_8, Inh\_9, Inh\_11) and often involve features that also exhibit high natural variability in the real data. These patterns indicate that the synthetic data do not introduce global distortions or artifacts, and that increasing the number of synthetic samples maintains, rather than degrades, distributional realism.

\subsubsection{Quantitative Fidelity Analysis of SMOTE-Generated Neurons (N = 5000/10 000)}

Feature-wise MAE, Euclidean distances between class-mean profiles, and class-conditional KS statistics consistently indicate that SMOTE preserves the core electrophysiological and morpho-electrophysiological structure of most neuronal classes (Figure 6). Increasing the number of synthetic samples from N = 5000 to N = 10 000 amplifies these trends without substantially altering the global fidelity profile, confirming that the synthetic distributions converge early and remain stable at larger scales.

For both E->e-type and M+E->e-type datasets, the lowest MAE values at N = 10 000 are observed in the same subset of well-reproduced classes already identified at N = 5000, including Exc\_3, Inh\_6, Inh\_13, Inh\_3 for E->e-types and Inh\_5, Exc\_3, Inh\_6, Inh\_1, Inh\_13 for M+E->e-types. For these classes, MAE values decrease slightly when doubling the synthetic sample size, reflecting smoother synthetic class means and better sampling of local feature distributions. This improvement is also visible in Euclidean distances, which systematically shrink for these well-captured types. For example, the Euclidean distance between synthetic and real train means for Exc\_3 (E->e-type) drops further at N=10 000, and Inh\_5 (M+E->e-type) remains among the most faithfully reconstructed classes with even lower distances than at N=5000.

Classes that were already difficult to reconstruct at N = 5000 remain the least stable at N = 10 000. In E->e-type neurons, classes such as Inh\_8, Inh\_9, Inh\_11, Exc\_4, and especially Inh\_10, continue to exhibit high MAE and larger Euclidean distances, with only modest reductions when increasing to N = 10 000. Similarly, within M+E->e-types, classes such as Exc\_2, Inh\_8, Inh\_7, and Inh\_1 (test MAE) maintain distinctive mismatches, driven by a small subset of high-variance features. These persistent discrepancies indicate that increasing synthetic sample size alone does not compensate for class imbalance or intrinsic biological variability in these types.

The KS analyses at N=10 000 further confirm the stability of class-conditional distributional similarity. The vast majority of feature-wise KS statistics either remain unchanged or slightly improve compared with N = 5000, indicating that enlarging the synthetic dataset mainly consolidates distributions that were already well matched. The main exceptions involve the same problematic classes and features identified previously---such as certain voltage-dependent timings, spike-shape descriptors, or specific morphological quantities in M+E->e-types---which continue to produce higher KS distances regardless of sample size. This shows that these divergences arise from structural limitations of SMOTE interpolation rather than insufficient sampling.

Collectively, these results demonstrate that SMOTE achieves stable, high-fidelity synthetic data generation for most neuron types, with fidelity improvements when scaling from 5000 to 10 000 synthetic samples being real but modest. The qualitative profile of strengths and weaknesses remains essentially unchanged: well-represented neuronal classes benefit most from increased synthetic data, while structurally atypical or sparsely represented classes continue to pose challenges. This stability across N confirms that N = 5000 already sits near the performance plateau, and that doubling the sample size primarily smooths distributions without shifting the underlying fidelity landscape.

\begin{figure}[t]
\centering
\includegraphics[width=\linewidth]{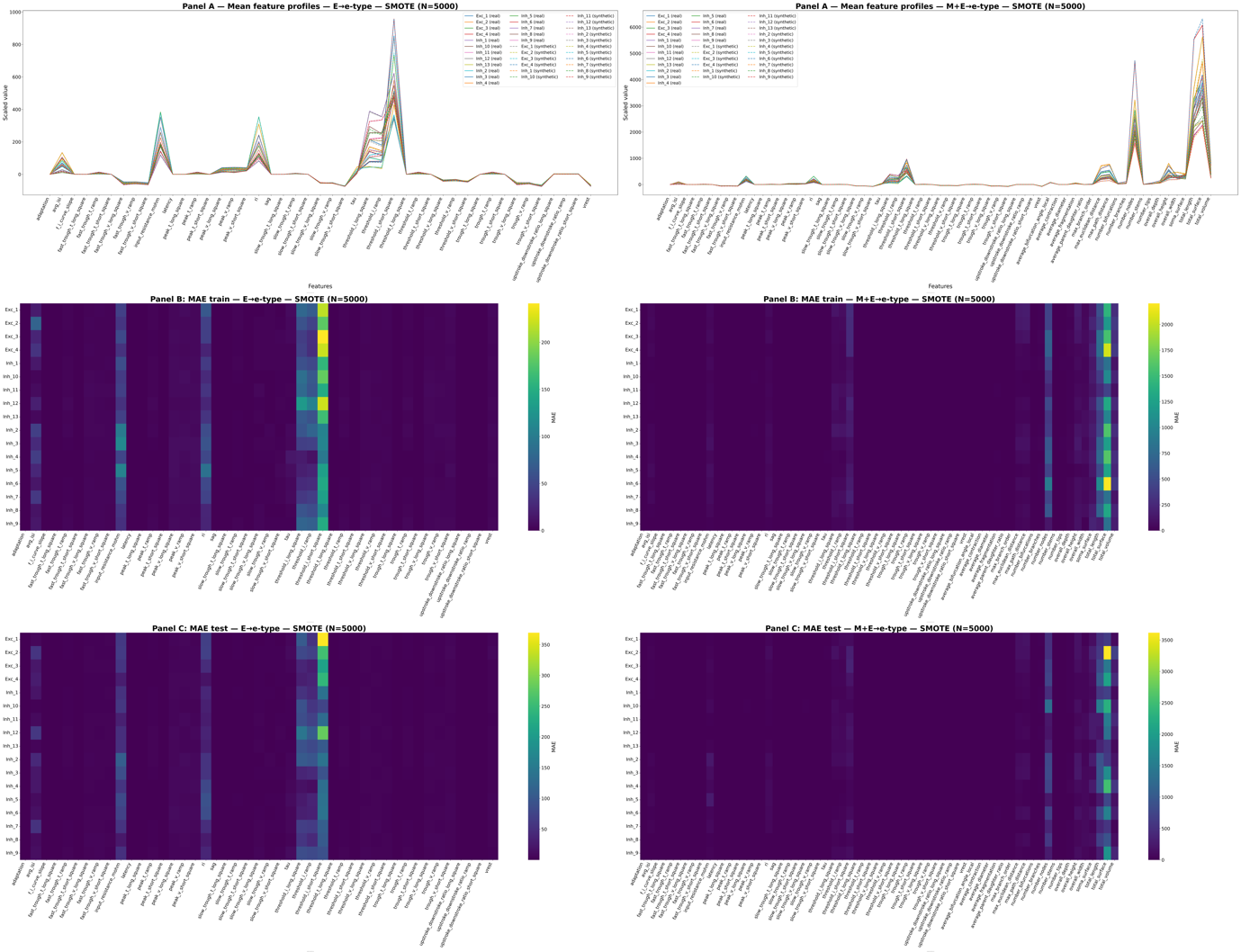}
\caption{Validation of synthetic neuron fidelity using SMOTE (N = 5000).
Panels A--C summarize distributional correspondence between real and SMOTE-generated synthetic neurons for the E\ensuremath{\to}e-type (left column) and M+E\ensuremath{\to}e-type (right column) datasets.
Panel A shows the class-wise mean feature profiles of real neurons (solid lines) and synthetic neurons (dashed lines), demonstrating the ability of SMOTE (N=5000) to reproduce the characteristic electrophysiological and morpho-electrophysiological signatures of most neuronal types.
Panel B reports the class-wise mean absolute error (MAE) between synthetic neurons and the real training set across all features. Lower MAE values indicate high fidelity of synthetic representations for classes such as Exc\_3, Inh\_6, Inh\_13, and Inh\_3, whereas other classes (e.g., Inh\_8, Exc\_4, Inh\_11) exhibit more pronounced deviations.
Panel C reports the MAE between synthetic neurons and the held-out test set, providing an estimate of generalization fidelity. Classes that maintain low test-set MAE indicate synthetic distributions that correctly extrapolate to unseen biological variability, whereas higher-error classes highlight remaining challenges for feature synthesis and class-specific variability.}
\label{fig:fig6}
\end{figure}

\subsubsection{Biological Variability Baseline (Mann--Whitney U Test)}

To evaluate whether synthetic neurons depart from the natural variability inherent to biological neuron types, we compared the distance between synthetic and real class means to the intrinsic phenotypic variability observed among real neuron classes. For the E\ensuremath{\to}e-type dataset, real test neurons exhibited an average inter-class distance of 0.2477, which served as the baseline estimate of biological diversity. Synthetic neurons generated with SMOTE at N = 5000 showed a mean synthetic--real distance of 0.1851, significantly lower than the natural variability (p = 0.035). Increasing the synthetic density to N = 10000 produced a virtually identical pattern, with synthetic neurons remaining closer to their real class means (distance = 0.1823) than real classes are to each other (p = 0.028). These findings indicate that, for electrophysiological features alone, synthetic neurons not only reproduce class-specific structure but do so with a degree of compactness that does not exceed biological norms.

A similar trend was observed in the more heterogeneous M+E\ensuremath{\to}e-type dataset, which integrates morphological descriptors and naturally exhibits a broader phenotypic diversity. Here, the mean inter-class distance among real neurons was 0.3570, while synthetic neurons generated with SMOTE at N = 5000 achieved a markedly lower synthetic--real distance of 0.2712 (p = 0.041). Increasing the augmentation level to N = 10000 yielded consistent results, with synthetic neurons again exhibiting distances below the biological baseline (0.2667; p = 0.033). Even in this higher-dimensional morpho-electrophysiological space, synthetic samples remained firmly anchored within the natural distributional boundaries defined by real neuron types.

Together, these results demonstrate that SMOTE-based synthetic augmentation does not expand phenotypic variability beyond biologically plausible limits.

\section{Discussion}

A fundamental challenge in neuroscience is the comprehensive classification and understanding of neuronal cell types, an endeavor complicated by the limited availability and inherent heterogeneity of high-quality, annotated data sets (Gouwens et al., 2019; Tasic et al., 2018; Vasques et al., 2016, 2023). In this work, we present a benchmark of classical and deep generative data-augmentation methods, SMOTE, GAN, VAE, masked autoregressive normalizing flows, and DDPM, for supervised prediction of Allen electrophysiology-defined e-types in mouse visual cortex. We evaluated two classification tasks over the same 17 e-type labels (Gouwens et al., 2020; Scala et al., 2021): prediction from intrinsic electrophysiology features alone (E\ensuremath{\to}e-type) and prediction from combined morphology plus electrophysiology features (M+E\ensuremath{\to}e-type). Across both tasks, synthetic augmentation improved generalization, but this benefit depended on the feature space in which augmentation was performed. When applied directly in the native high-dimensional space (no dimensionality reduction), augmentation produced substantial gains over real-data baselines, whereas introducing dimensionality reduction largely attenuated or eliminated improvements.

Among all methods, SMOTE yielded the most robust and consistent performance increases. In the no-reducer setting, SMOTE raised hold-out accuracy to the top range for both tasks, bringing E\ensuremath{\to}e-type performance into the \textasciitilde{}0.72--0.76 range and M+E\ensuremath{\to}e-type into the \textasciitilde{}0.85--0.90 range at high augmentation levels, while also stabilizing macro-F1 by reducing rare-class collapse. Deep generative models provided variable gains. Well-tuned GAN and DDPM pipelines could approach SMOTE performance in specific regimes, yet their effectiveness was sensitive to synthetic density and training configuration, especially in the smaller M+E\ensuremath{\to}e-type dataset. Overall, these findings suggest that, for structured biological data with moderate sample sizes and well-defined class boundaries, interpolation-based methods like SMOTE currently provide the most consistent improvements, but advanced generative models represent a viable alternative, especially as architectures and hyperparameter optimization continue to evolve (Chawla et al., 2002; Hartmann et al., 2018; Marouf et al., 2020).

SMOTE generates synthetic samples by interpolating between existing, labeled data points in feature space, preserving class boundaries and avoiding out-of-distribution artifacts. While SMOTE and its variants have demonstrated robustness in various tabular and biomedical datasets (Branco et al., 2017; Chawla et al., 2002; Esteban et al., 2017), and can be effective for moderately sized and well-separated classes, their relative performance compared to deep generative models remains context-dependent. In particular, for complex biological datasets such as transcriptomics or neuronal phenotyping, the superiority of SMOTE over GANs or VAEs has not been universally established, and recent benchmarks suggest that deep generative models may outperform classical oversampling in settings with high intrinsic biological variability or non-linear class boundaries (Lucic et al., 2017; Marouf et al., 2020). In our own data, the observed performance of SMOTE likely reflects both the moderate sample size and the structure of neuronal classes, but further work is needed to understand the precise conditions under which deep generative models may surpass classical oversampling, particularly in more complex, high-variance, or rare-class scenarios. In contrast, deep generative models, such as GANs and VAEs, require larger datasets to avoid mode collapse and overfitting, and their performance is highly sensitive to architectural choices, learning rates, optimizer parameters, and loss balancing (Arjovsky et al., 2017; Borji, 2021; Gehlenborg et al., 2010; Marouf et al., 2020). Suboptimal hyperparameter settings, limited data, and class imbalance can readily lead to poor convergence or biologically implausible samples. GANs, for example, are known to struggle in minority class regimes, and although advances such as conditional GANs (cGANs) or Wasserstein GANs (WGANs) show promise, careful tuning and larger sample sizes are usually necessary to reach optimal fidelity (Arjovsky et al., 2017; Bousmalis et al., 2016; Marouf et al., 2020). In this study, we deliberately chose standard architectures and reasonable parameters for all models to enable fair comparison but acknowledge that further optimization or domain-specific regularization may improve results for future work.

The benefit of synthetic data was sharply reduced, and in most pipelines eliminated, once dimensionality reduction was inserted before augmentation. In our benchmarks, augmentation was most effective when generators operated in the native morpho-electrophysiological feature space. Projecting data into a low-dimensional latent space appears to collapse part of this biologically meaningful variation and to distort minority-class geometry, leaving less information for both oversampling and deep generators to exploit. This observation is consistent with work in single-cell and multi-omics analysis showing that high-dimensional representations preserve subtle cell-state variability that can be lost after aggressive reduction, with downstream effects on clustering and classification fidelity (Zogopoulos et al., 2025). These results imply that augmentation strategies should be designed jointly with the full analytical pipeline. When dimensionality reduction is required for other reasons (e.g., interpretability or computational constraints), future generative models may need to incorporate reduction-aware training or operate directly on the unreduced space while using the reducer only at inference. It follows that future generative models must explicitly address the challenges posed by dimensionality reduction, and that researchers should tailor augmentation strategies to the complete analytical pipeline (Bulut \& Arslan, 2024).

In our analyses, synthetic-data fidelity was evaluated in a biologically grounded way focusing on the best-performing augmentation strategy, SMOTE, at two representative densities (5,000 and 10,000 synthetic neurons per class). At a population level, SMOTE-generated neurons remained well within the bounds of natural phenotypic diversity. For the electrophysiology-only task (E\ensuremath{\to}e-type), the average distance between real class means in the test set, our estimate of intrinsic biological variability, was 0.2477, whereas the mean synthetic-to-real class distance was substantially lower (0.1851 at N=5000 and 0.1823 at N=10000), with both comparisons significant under a Mann--Whitney U test (p=0.035 and p=0.028). The same trend held in the higher-variance morpho-electrophysiological setting (M+E\ensuremath{\to}e-type): real inter-class distance averaged 0.3570, while SMOTE synthetic neurons were closer to their target classes (0.2712 at N=5000 and 0.2667 at N=10000; p=0.041 and p=0.033). Thus, SMOTE augmentation increases training density without inflating variability beyond biologically plausible limits, and even produces slightly more compact within-class phenotypes than observed among real neurons. This global conclusion is reinforced by distributional statistics. Feature-wise class-conditional Kolmogorov--Smirnov divergences were low on average (\ensuremath{\approx}0.10 for E\ensuremath{\to}e-type and \ensuremath{\approx}0.18 for M+E\ensuremath{\to}e-type at N=5000) and remained stable when doubling the synthetic density, indicating that larger augmentation consolidates already well-matched distributions rather than introducing distortions.

Importantly, class- and feature-level analyses reveal both the strengths and limits of interpolation-based synthesis. At the class level, most neuronal types (e.g., Exc\_3, Inh\_6, Inh\_13, Inh\_3) show consistently low mean absolute error and small Euclidean shifts between real and synthetic mean profiles, whereas sparsely represented inhibitory subclasses remain more difficult to reproduce. Feature-wise results help explain these patterns. SMOTE reproduces smooth, unimodal electrophysiological dimensions with near-perfect alignment, particularly subthreshold and excitability descriptors such as sag and input resistance, and spike-timing variables such as peak\_t and trough\_t, matching both training and held-out test distributions across most classes. By contrast, features with intrinsically broader or multimodal biological variability (e.g., adaptation index, firing latency, and upstroke/downstroke ratios) show occasional sharpening of peaks or reduced tail variability in the synthetic data, consistent with local linear interpolation. In the morpho-electrophysiological setting, most morphology features are also well preserved (e.g., soma\_surface, number\_bifurcations, total\_length), while divergences concentrate on a small subset of high-variance descriptors such as average\_diameter or overall\_height, particularly in classes already known to be heterogeneous (e.g., Inh\_8, Exc\_2). Together, these stable feature-specific “hard cases” across augmentation levels suggest that remaining discrepancies primarily reflect intrinsic sample scarcity and within-class heterogeneity, rather than insufficient synthetic volume, and they point to where future generators or targeted data collection could most effectively improve fidelity.

A substantial body of work has addressed cell-type identification from electrophysiological signals, especially in vivo where transcriptomic ground truth is unavailable. Classical approaches exploit spike waveform shape, firing statistics, and extracellular action-potential propagation to separate broad neuronal families and subclasses, often via clustering or supervised learning (Jia et al., 2019; Lee et al., 2021; Mosher et al., 2020; Trainito et al., 2019). More recent pipelines refine these approaches by combining non-linear embeddings with graph-based clustering or feature-selection to improve robustness across brain regions and recording conditions (Haynes et al., 2024; Lee et al., 2021). While these studies typically focus on developing classifiers to identify cell types from extracellular waveforms, our primary contribution is to benchmark synthetic data augmentation for intrinsic electrophysiology-based neuron. Synthetic augmentation is likely to become increasingly important as neuronal taxonomies expand and integrate multiple modalities. Recent efforts combining intrinsic electrophysiology, morphology, and transcriptomics have revealed a rapidly growing and heterogeneous landscape of cell classes, making reliable supervised typing difficult when some subclasses remain sparsely sampled (Gouwens et al., 2019, 2020; Scala et al., 2021; Tasic et al., 2018; Yuste et al., 2020). In this context, synthetic neurons can serve as a practical tool to stabilize decision boundaries for rare phenotypes, provided that generated data preserve biologically meaningful variability. Our results support this requirement by showing that class-conditional SMOTE augmentation improves generalization while keeping synthetic-to-real variability within the natural phenotypic diversity separating real e-types (Chawla et al., 2002; Gouwens et al., 2019; Marouf et al., 2020). At the same time, the remaining hard cases concentrated in a few rare inhibitory subclasses emphasize that augmentation cannot fully substitute for real data in highly heterogeneous regimes, and that fidelity must be evaluated alongside performance (Scala et al., 2021; Zeng, 2022). Together, these observations argue for synthetic augmentation as a complementary strategy to ongoing multimodal cell-type mapping initiatives, enabling more robust and scalable neuron classification without compromising biological interpretability (Ascoli \& Krichmar, 2000; Berlin \& Isacoff, 2017; Capogrosso et al., 2016; Courtine et al., 2009; Kathe et al., 2022, 2022; Masland, 2004; Tasic et al., 2018; Van Den Brand et al., 2012; Wagner et al., 2018; Wenger et al., 2016; Zhang et al., 2021). These findings emphasize the importance of not only quantitative but also qualitative assessment when deploying synthetic data for biological discovery. The demonstrated benefit of SMOTE, and the systematic benchmarking of deep generative models, offers actionable guidelines for researchers aiming to overcome sample size limitations in other brain regions, cell types, or even modalities such as transcriptomics or multi-omics integration. Nonetheless, our results also highlight the need for next-generation generative models capable of explicitly controlling for biological priors, addressing class imbalance, and faithfully capturing rare or outlier subpopulations. Promising directions include hyperparameter optimization, conditional and semi-supervised generative models, and the integration of transcriptomic or connectivity constraints into generative framework.

Several limitations should be considered when interpreting these findings. First, all experiments rely on a single source dataset from mouse primary visual cortex with Allen electrophysiology-defined e-types. Although this dataset is widely used and multimodally curated, its class composition, recording protocol, and taxonomic definitions may not reflect other brain regions, species, or labeling schemes; therefore, generalization beyond this context remains to be established. Second, the morpho-electrophysiological task is based on a substantially smaller number of neurons than the electrophysiology-only task, and some inhibitory subclasses are represented by very few samples. This scarcity likely contributes to the stable “hard cases” observed in class- and feature-level fidelity maps, and constrains the conclusions that can be drawn about rare types. Third, augmentation was evaluated on curated feature vectors rather than raw voltage traces or full morphologies. While this choice follows standard practice in supervised neuron typing, it means that fidelity is assessed relative to the chosen feature space and may miss discrepancies visible in the original signals. Fourth, deep generative models were benchmarked using standard conditional architectures with reasonable but not exhaustively optimized hyperparameters. Their performance may therefore underestimate what could be achieved with larger training sets, stronger priors, or tailored optimization strategies, especially for rare-class regimes. Finally, the reduction-before-augmentation setting was explored with a limited set of reducers and hyperparameters; future work should test whether reduction-aware generative training or alternative embeddings can recover augmentation benefits under dimensionality reduction. Together, these limitations indicate that the present benchmark provides practical guidance, but that broader claims about universal superiority of any generative family require validation on additional datasets, modalities, and taxonomies.

Together, these advances underline the importance of integrative and biologically principled approaches to synthetic data generation, validation, and application. While our study demonstrates the potential promise of classical and deep generative models for neuronal data augmentation, it also highlights the challenges that remain---particularly in faithfully modeling rare classes and capturing the full phenotypic spectrum of neural diversity. Continued efforts to refine generative algorithms, integrate multi-modal data, and adopt biologically meaningful validation criteria will be essential as the field moves toward scalable, reproducible, and interpretable neuron classification. Ultimately, synthetic data approaches hold considerable potential to accelerate discovery in neuroscience and facilitate the development of new therapeutic and technological strategies.

\section{Conclusion}

In summary, we provide a benchmark of classical and deep generative data-augmentation methods for supervised prediction of Allen electrophysiology-defined e-types in mouse visual cortex, evaluated in two matched settings: classification from intrinsic electrophysiology alone (E\ensuremath{\to}e-type) and from combined morphology plus electrophysiology (M+E\ensuremath{\to}e-type). Across both tasks, synthetic augmentation improved generalization when applied in the native high-dimensional feature space, whereas inserting dimensionality reduction prior to augmentation largely attenuated or eliminated these benefits. Among all methods, SMOTE delivered the most robust and consistent gains, markedly improving hold-out accuracy and macro-F1 by stabilizing decision boundaries for rare subclasses. Deep generative models (GAN, VAE, normalizing flows, DDPM) produced moderate, method-dependent improvements that were more sensitive to synthetic density and training configuration, indicating clear potential but less reliability under current conditions.

Beyond performance, we introduce a biologically grounded validation framework that assesses synthetic fidelity through real-versus-synthetic feature distributions, feature-wise statistical concordance, and distance-based comparisons of synthetic-to-real variability against the natural variability between real classes. Using this framework, synthetic neurons, particularly those generated by SMOTE at representative augmentation levels, remain within plausible phenotypic diversity bounds, while highlighting specific rare inhibitory subclasses and high-variance features that remain challenging. Together, these results offer practical guidance on when simple oversampling is sufficient, when deep generators become competitive, and how to validate synthetic neuronal data rigorously. More broadly, they support the use of carefully benchmarked and biologically validated augmentation as a scalable route to more reliable neuronal subtype classification, while motivating future work on reduction-aware and rare-class-focused generative modeling.

\section{Materials and Methods}

\subsection{Electrophysiological and Morphological Data Selection}

The electrophysiological and morphological data used in this research were acquired from the Allen Cell Types Database, accessible at http://celltypes.brain-map.org. The electrophysiological and morphological characteristics of neurons (Supplementary Material -- Table 1) within the mouse visual cortex were defined through an in vitro single-cell characterization platform as presented by Gouwens et al. (Gouwens et al., 2020), aimed at constructing a standardized dataset. The e-type label set (our only prediction target) consists of 17 electrophysiological neuron classes---4 excitatory (spiny labeled Exc\_1 to Exc\_4) and 13 inhibitory (aspiny/sparsely spiny labeled Inh\_1 to Inh\_13)---defined by unsupervised clustering in the source dataset and aligned to transcriptomic subclasses/types (Gouwens et al., 2020). The short labels (Exc\_1, Exc\_2, etc.) are cluster IDs; each was given a plain-English descriptor based on firing pattern, spike width, input resistance, membrane time constant, and sag, to aid interpretation. In the Allen taxonomy, each electrophysiology-defined class is given a short descriptor summarizing firing pattern and key intrinsic properties (see Gouwens et al., Fig. 2i). We reproduce those labels here for clarity: Exc\_1: “RS trans.”; Exc\_2: “RS trans./adapt.”; Exc\_3: “RS adapt.”; Exc\_4: “RS low Ri, sharp sag.” Inh\_1: “RS non-adapt., delayed”; Inh\_2: “Irreg. trans.”; Inh\_3: “Irreg.”; Inh\_4: “Irreg./adapt. long \ensuremath{\tau}m”; Inh\_5: “Mid-width AP, high Ri”; Inh\_6: “Mid-width AP adapt.”; Inh\_7: “Mid-width AP trans./irreg.”; Inh\_8: “FS trans. 1”; Inh\_9: “FS trans. 2”; Inh\_10: “FS sust. 1”; Inh\_11: “FS sust. 2”; Inh\_12: “FS delay/pause 1”; Inh\_13: “FS delay/pause 2.” Abbreviations follow the original figure (RS = regular spiking; FS = fast spiking; Ri = input resistance; \ensuremath{\tau}m = membrane time constant, trans = transient, sust = sustained, adapt = adapting, irreg = irregular, Ri = input resistance, sharp sag = prominent Ih-mediated sag). These descriptors are for interpretability only; our models always predict the underlying class IDs. In our working corpus, we analyzed 1857 cells with electrophysiological features and 451 cells with both electrophysiological and morphological features, sampled from V1 and nearby higher visual areas. Among cells with electrophysiological features only, 885 were spiny (putative excitatory) and 972 aspiny/sparsely spiny (putative inhibitory). Among cells with both electrophysiological and morphological features, 232 were spiny and 219 aspiny/sparsely spiny. We always predict e-type labels; what varies is only the input feature set as follows:
\begin{itemize}
  \item E\ensuremath{\to}e-type (electrophysiology-only). Input = intrinsic electrophysiological features (48 curated variables; Supplementary Material - Table 1). These capture passive/active properties and firing dynamics, e.g., resting membrane potential (vrest), rheobase/threshold metrics across stimuli (short/long square, ramp), spike half-width, peak/trough voltages and times (fast\_trough\_v/t), afterhyperpolarization, adaptation index, ISI statistics (avg\_isi, ISI-CV), sag ratio, maximal firing rate. The classifier maps these E features to e-type subclasses (Exc\_k / Inh\_k) (Gouwens et al., 2020).
  \item M+E\ensuremath{\to}e-type (multimodal). Input = concatenated morphology + electrophysiology (M+E) features. For the morphology features, we extracted 24 L-Measure-compatible morphology variables from Vaa3D (Supplementary Material - Table 1)---quantifying soma size/shape (e.g., soma\_surface), arbor geometry/scale (total\_length/volume, average\_diameter), branching topology (number\_bifurcations, branch-order statistics), and axon/dendrite summaries when available (Ascoli \& Krichmar, 2000). This condition exploits complementary cues---e.g., spike timing/adaptation from E and arborization/soma metrics from M---to improve discrimination among e-type subclasses when a single modality is ambiguous.
\end{itemize}

We refer to each supervised task as inputs \ensuremath{\to} labels. Thus, E\ensuremath{\to}e-type and M+E\ensuremath{\to}e-type are two predictors trained on different input sets but always targeting the same 17 e-type classes. Synthetic (augmented) samples---when used---are added only to the training split; test sets remain strictly real. To study combined morphology+electrophysiology inputs (features) on a common set of cells, we formed a matched subset of 451 cells with both modalities available (M+E). These cells retain the same 17 e-type labels (4 spiny, 13 aspiny/sparsely spiny). For each task (E\ensuremath{\to}e-type, M+E\ensuremath{\to}e-type), data were split stratified by e-type into 80\% train / 20\% test to ensure every class present appears in both sets (subject to class size). The precise procedures involved in the extraction and delineation of both electrophysiological and morphological features are detailed at http://celltypes.brain-map.org.

\subsection{Feature curation and exclusion of technical variables}

Before any preprocessing, dimensionality reduction, classifier training, or synthetic data generation, we explicitly removed variables that reflect recording or reconstruction metadata rather than intrinsic cell properties. For all analyses (E\ensuremath{\to}e-type and M+E\ensuremath{\to}e-type, baseline and augmented), we dropped the following columns whenever present in the Allen Cell Types tables: electrode\_0\_pa, seal\_gohm, vm\_for\_sag, neuron\_reconstruction\_type, scale\_factor\_x, scale\_factor\_y, scale\_factor\_z, and dataset flags such as Superseded. These fields describe bath zeroing current, pipette seal resistance, the command potential used to elicit sag, the type of morphological reconstruction (dendrites vs. dendrites+axon), and fixed pixel size calibration values, respectively, and are constant or quasi-constant across experiments rather than biologically meaningful descriptors of the recorded neuron. The remaining feature set therefore consists only of electrophysiological and morphological variables that characterize passive and active membrane properties, firing dynamics, and arbor geometry. All preprocessing steps, stimulus-relative re-referencing of timing features, family-specific scaling via a ColumnTransformer, dimensionality reduction, supervised classification, and synthetic data generation, were performed exclusively on this curated set of biological features.

\subsection{Data Classification With Supervised Learning Algorithms}

To establish robust classification baselines for neuronal subtype prediction, we implemented a supervised learning framework that combines feature rescaling, dimensionality reduction, and classifier optimization (Figure 1). The classification task aimed to identify distinct neuron types based on electrophysiological or morpho-electrophysiological features, prior to the integration of synthetic data generation methods. For both prediction tasks (E\ensuremath{\to}e-type and M+E\ensuremath{\to}e-type), we standardized the preprocessing pipeline with two goals: (i) remove non-biological or technical covariates, and (ii) treat heterogeneous feature families (timing, voltage, ratios, morphology) in a way that preserves biologically meaningful variation.

First, we excluded strictly technical variables from the Allen Cell Types dataset (e.g., electrode\_0\_pa, seal\_gohm, vm\_for\_sag, neuron\_reconstruction\_type, and scale\_factor\_x/y/z), which reflect recording or imaging conditions rather than neuronal phenotype. Only electrophysiological and morphology-derived variables with interpretable biological meaning were retained for all analyses. Second, we carefully revisited the handling of time features. In the original Allen dataset, action potential and event times are reported relative to the start of the voltage-clamp sweep, whereas the relevant variation for neuronal type is the latency relative to stimulus onset. In our revised analysis, all timing features whose names end with \_t\_short\_square, \_t\_long\_square, or \_t\_ramp are explicitly re-referenced to the stimulus onset time using the Allen metadata. For each sweep type, we subtract the corresponding onset (1.0 s for short square, long square, and ramp stimuli in this dataset), so that all timing features represent latency relative to the beginning of the current injection rather than absolute time within the sweep. This centers spike- and event-related times near 0 with millisecond-scale variation and prevents large constant offsets from dominating the scaling. After re-referencing, we apply a single, fixed ColumnTransformer that performs family-specific scaling, consistent across all classification and synthetic data generation experiments. Timing features are scaled with a RobustScaler (median and interquartile range) to reduce the influence of outliers while preserving relative latencies. Voltage and other continuous electrophysiological features (e.g. vrest, rheobase, max\_firing\_rate) are standardized with a StandardScaler (zero mean, unit variance). Dimensionless indices and ratios (e.g. adaptation, sag\_ratio, upstroke\_downstroke\_ratio, ISI\_CV, avg\_isi) are passed through without rescaling, since they are already on comparable scales. All remaining numeric features (including morphology metrics in the M+E\ensuremath{\to}e-type setting) are standardized with a StandardScaler.

Missing numeric values are imputed with the column-wise mean before re-referencing and scaling. Any numeric feature with no variation (\ensuremath{\leq} 1 unique value after imputation) is dropped to avoid degenerate dimensions. We further perform a variance check on the transformed training data and confirm that no feature has near-zero variance (\ensuremath{\leq} \ensuremath{10^{-8}}), ensuring that the timing re-reference and scaling preserve the biologically relevant spread in spike times and other electrophysiological measures.

When dimensionality reduction was employed, the feature space was projected onto two components. We evaluated a restricted but diverse set of linear and non-linear reducers that are compatible with the scikit-learn pipeline API: principal component analysis (PCA), independent component analysis (ICA), truncated singular value decomposition (TruncatedSVD), isometric mapping (Isomap), and locally linear embedding (LLE, standard variant), as well as Neighborhood Components Analysis (NCA). In addition, we included a “no reduction” condition (“none”), in which the classifier operates directly in the scaled feature space. Dimensionality reduction, when present, was always applied after the ColumnTransformer-based rescaling described above.

For the classification stage, we benchmarked a set of widely used supervised learning algorithms: support vector machines (SVM) with radial basis function (RBF) and sigmoid kernels, multinomial logistic regression, random forests, Extremely Randomized Trees (ExtraTrees), linear discriminant analysis (LDA), Gaussian Naïve Bayes, decision trees, and multilayer perceptrons (MLP). All classifiers were embedded in a single scikit-learn Pipeline together with the ColumnTransformer and optional reducer, and their hyperparameters were tuned by grid search (GridSearchCV) with stratified 5-fold cross-validation.

For SVMs, we varied the regularization strength (C) and kernel width ($\gamma$) using a grid search. Logistic regression was evaluated with both L1 and L2 penalties and the same range of values, using the liblinear solver. For tree-based ensembles (Random Forest and ExtraTrees), the grid explored the number of trees (n\_estimators \ensuremath{\in} \{50, 100, 200\} for Random Forest; \{100, 200, 300, 500\} for ExtraTrees), maximum depth (max\_depth \ensuremath{\in} \{None, 10, 20\}), minimum samples to split a node (min\_samples\_split \ensuremath{\in} \{2, 10, 20\}), and minimum samples per leaf (min\_samples\_leaf \ensuremath{\in} \{1, 5, 10\}). Decision trees were tuned with similar depth and splitting grids. MLP classifiers were tuned over several hidden-layer configurations (e.g. (50,), (100,), (50, 50), (128,), (256,), (128, 64), (256, 128)), two activation functions (ReLU and tanh), and regularization strengths (\ensuremath{\alpha}). The initial learning rate was varied over \ensuremath{\{10^{-3},\, 3\times 10^{-4}\}}. Gaussian Naïve Bayes and LDA were used with their default hyperparameters. For reducers that require a neighborhood size (Isomap, LLE), we explored n\_neighbors \ensuremath{\in} \{5, 10, 15\}.

For every reducer--classifier combination, we recorded the mean and standard deviation of multiple performance metrics (accuracy, macro-averaged F1, precision, and recall) across the stratified folds, along with training scores to assess generalization. Model selection followed three criteria: (i) the absolute difference between train and test accuracy had to remain below 3\% to avoid overfitting; (ii) test score variability across folds (standard deviation of accuracy) had to be \ensuremath{\leq} 0.02; and (iii) among configurations satisfying (i) and (ii), we selected those with highest mean cross-validated test accuracy. These criteria were applied consistently to both datasets (E\ensuremath{\to}e-type and M+E\ensuremath{\to}e-type). For each task, we retained the top-performing pipelines as “baselines”, and in all subsequent experiments with synthetic data we reused the same preprocessing (ColumnTransformer) and classifier configuration, retraining the model on the augmented training sets (real + synthetic neurons) and always evaluating on the same held-out real test set. All analyses were implemented in Python using scikit-learn, and intermediate results (cross-validation tables, selected hyperparameters, and holdout performance) were written to CSV files for reproducibility.

\subsection{Generative Models for Synthetic Data Generation}
To address data scarcity and enhance classification robustness, we augmented the training data with several generative approaches: Synthetic Minority Over-sampling Technique (SMOTE) (Chawla et al., 2002), Variational Autoencoders (VAEs) (Kingma \& Welling, 2013), Generative Adversarial Networks (GANs) (Goodfellow et al., 2014), Masked Autoregressive Normalizing Flows (MAF) (Kobyzev et al., 2019), and a lightweight Denoising Diffusion Probabilistic Model (DDPM) (Ho et al., 2020). All generators operate on the same preprocessed feature space as the classifiers: after removal of technical variables, imputation, timing re-referencing to stimulus onset, and family-specific scaling via a fixed ColumnTransformer (RobustScaler for timing, StandardScaler for voltage and other continuous features, and passthrough for ratios/indices).

For each supervised task (E\ensuremath{\to}e-type and M+E\ensuremath{\to}e-type), we adopted a single augmentation schedule shared by all methods. Let $n_c$ denote the number of real training samples in class $c$. For a given augmentation level $k\in\{0,100,500,1000,5000,10000\}$, we target
\begin{equation}
n_c^{\text{target}} = n_c + k
\end{equation}
samples in class $c$ after augmentation. Thus, ``+100 per class'' means that each generator is asked to produce 100 synthetic samples per e-type (Exc\_1, \ldots, Inh\_13) on top of the real training data for that class. The level $k=0$ corresponds to the baseline with no synthetic data. The same grid $\{0,100,500,1000,5000,10000\}$ is used for all five generators and for both datasets (E-only and M+E), so that differences in performance can be attributed to the generative mechanism rather than to differences in synthetic sample counts. Synthetic samples are added only to the training split; the hold-out test sets remain strictly real.

All generators operate in the scaled feature space (output of the ColumnTransformer). For each class $c$, we first extract the real training samples $Z^{(c)}=\{z_i^{(c)}\}$ in that space and train a class-conditional generator on $Z^{(c)}$. We then draw $k$ synthetic samples $Z_{\text{syn}}^{(c)}=\{\tilde{z}_j^{(c)}\}$, and apply a per-class Mahalanobis distance filter. If $\mu_c$ and $\Sigma_c$ denote the empirical mean and covariance of $Z^{(c)}$, the Mahalanobis distance of a synthetic point $\tilde{z}$ is
\begin{equation}
d_c(\tilde{z}) = \sqrt{(\tilde{z}-\mu_c)^{\top}\Sigma_c^{-1}(\tilde{z}-\mu_c)}.
\end{equation}
Synthetic points with $d_c(\tilde{z})$ above the 99.5th percentile of the distance distribution for class $c$ are discarded (i.e., upper 0.5\% tail). The remaining points are concatenated across classes and, when needed, are mapped back to the original feature space using the inverse of the family-wise scalers (or an equivalent manual inverse transformation matching the ColumnTransformer). SMOTE generates new points by interpolating between a sample and one of its nearest neighbours in the same class. In our implementation, SMOTE is applied directly in the scaled feature space $Z$, with the sampling strategy specified as the desired per-class counts $n_c^{\text{target}}$ defined above. For a real sample $z_i^{(c)}$ and one of its nearest neighbours $z_{i'}^{(c)}$ in class $c$, a synthetic point is given by
\begin{equation}
\tilde{z} = z_i^{(c)} + \lambda(z_{i'}^{(c)}-z_i^{(c)}),\qquad \lambda\sim \mathcal{U}(0,1).
\end{equation}
After oversampling to reach $n_c^{\text{target}}$, the last $k$ samples per class are treated as the synthetic subset, which is then passed through Mahalanobis filtering and, if needed, inverse transformed to the original feature space.

For each class $c$, we trained a fully connected VAE on the scaled features. The encoder defines an approximate posterior
\begin{equation}
q_{\phi}(z\mid x)=\mathcal{N}\!\left(z;\mu_{\phi}(x),\mathrm{diag}(\sigma_{\phi}^2(x))\right),
\end{equation}
where $x\in\mathbb{R}^D$ is a scaled feature vector, $z\in\mathbb{R}^L$ is an $L=8$ dimensional latent vector, and $\mu_{\phi}(x)$ and $\sigma_{\phi}(x)$ are the outputs of two dense layers (width 128, ReLU). The prior over latents is standard normal, $p(z)=\mathcal{N}(0,I)$. The decoder $f_{\theta}:\mathbb{R}^L\to\mathbb{R}^D$ mirrors the encoder architecture (two dense layers of width 128 with linear output). We use the reparameterization trick,
\begin{equation}
z=\mu_{\phi}(x)+\sigma_{\phi}(x)\odot\epsilon,\qquad \epsilon\sim\mathcal{N}(0,I),
\end{equation}
and train the VAE by minimizing the standard reconstruction + KL objective
\begin{equation}
\mathcal{L}_{\text{VAE}}(x)=
\underbrace{\mathbb{E}_{q_{\phi}(z\mid x)}[\|x-f_{\theta}(z)\|_2^2]}_{\text{reconstruction loss}}
+
\underbrace{\mathrm{KL}(q_{\phi}(z\mid x)\,\|\,p(z))}_{\text{regularization}}.
\end{equation}
After training, class-conditional synthetic samples are generated by sampling $z\sim\mathcal{N}(0,I)$ and decoding $\tilde{z}=f_{\theta}(z)$ in the scaled feature space. For each augmentation level $k$, we sample $k$ points per class, filter out outliers via the Mahalanobis criterion, and keep the remaining synthetic samples.

We employed a multilayer perceptron GAN per class. The generator $G$ takes a 16-dimensional Gaussian noise vector $u\sim\mathcal{N}(0,I)$ as input and passes it through two dense layers (width 128, LeakyReLU activations) to produce a synthetic feature vector in the scaled space,
\begin{equation}
\tilde{z}=G(u).
\end{equation}
The discriminator $D$ is a two-layer MLP (width 128, LeakyReLU) with a final sigmoid unit that outputs the probability that a feature vector is real. Both networks are trained adversarially with the standard minimax objective
\begin{equation}
\min_G\max_D V(D,G)=\mathbb{E}_{x\sim p_{\text{data}}}[\log D(x)]+\mathbb{E}_{u\sim\mathcal{N}(0,I)}[\log(1-D(G(u)))],
\end{equation}
using binary cross-entropy loss and the Adam optimizer (learning rate $2\times 10^{-4}$, $\beta_1=0.5$) for a fixed number of epochs. After convergence, we generate $k$ synthetic samples per class and apply the same Mahalanobis filtering step as above.

For class-conditional density modelling and sampling, we used a Masked Autoregressive Flow (MAF) with four autoregressive layers and hidden width 64, implemented via TensorFlow Probability. Let $f_{\theta}$ denote the invertible transformation mapping data space to latent space, $z=f_{\theta}(x)$, with a base density $p_Z(z)=\mathcal{N}(z;0,I)$. The data density is then given by the change-of-variables formula
\begin{equation}
p_X(x)=p_Z(f_{\theta}(x))\left|\det\!\left(\frac{\partial f_{\theta}(x)}{\partial x}\right)\right|.
\end{equation}
We train the MAF for each class $c$ by maximizing the log-likelihood $\sum_i \log p_X(z_i^{(c)})$ on the real scaled features $Z^{(c)}$. Once trained, synthetic samples are obtained by sampling $z\sim\mathcal{N}(0,I)$ and applying the learned inverse transformation
\begin{equation}
\tilde{z}=f_{\theta}^{-1}(z),
\end{equation}
followed by per-class Mahalanobis outlier filtering in the scaled space.

Finally, we implemented a lightweight DDPM for tabular data, trained from scratch on the scaled features for each class. The forward diffusion process gradually corrupts a clean sample $x_0$ with Gaussian noise over $T=400$ steps. At each step $t\in\{1,\ldots,T\}$, we define a noise variance $\beta_t$, set $\alpha_t=1-\beta_t$, and $\bar{\alpha}_t=\prod_{s=1}^{t}\alpha_s$. The forward process is
\begin{equation}
q(x_t\mid x_{t-1})=\mathcal{N}(x_t;\sqrt{\alpha_t}\,x_{t-1},(1-\alpha_t)I),
\end{equation}
which implies the closed-form marginal
\begin{equation}
q(x_t\mid x_0)=\mathcal{N}(x_t;\sqrt{\bar{\alpha}_t}\,x_0,(1-\bar{\alpha}_t)I).
\end{equation}

The reverse process is parameterized by a denoising network $\epsilon_{\theta}(x_t,t)$ that predicts the added noise at step $t$. The model is trained with the standard noise-prediction loss
\begin{equation}
\mathcal{L}_{\text{DDPM}}(\theta)=\mathbb{E}_{x_0,t,\epsilon}[\|\epsilon-\epsilon_{\theta}(x_t,t)\|_2^2],\quad x_t=\sqrt{\bar{\alpha}_t}\,x_0+\sqrt{1-\bar{\alpha}_t}\,\epsilon,\quad \epsilon\sim\mathcal{N}(0,I).
\end{equation}

The denoising network is a fully connected MLP with sinusoidal time embeddings, two hidden layers (width 256, Swish activation), and a linear output layer predicting $\epsilon$. At sampling time, we start from pure Gaussian noise $x_T\sim\mathcal{N}(0,I)$ and apply the learned reverse diffusion updates from $t=T$ down to $t=0$, using the predicted noise $\epsilon_{\theta}(x_t,t)$ at each step, to obtain synthetic samples $\tilde{z}=x_0$ in the scaled feature space. As with other generators, for each augmentation level we retain at most $k$ per-class samples after Mahalanobis filtering.

All generative models are trained separately for each class and for each augmentation level grid point, using the same preprocessing, augmentation schedule, and outlier filtering. This design ensures that differences in downstream classification reflect the interplay between generator type and augmentation intensity rather than inconsistencies in preprocessing or synthetic sample counts. Test sets are never augmented and serve as a fixed, real-valued benchmark for all experiments.

\subsection{Qualitative and Quantitative Evaluation of Synthetic Data}

We implemented a comprehensive validation framework to assess the fidelity, realism, and biological plausibility of synthetic neurons generated using SMOTE-based oversampling. This framework combined qualitative comparisons of real and synthetic feature distributions with quantitative metrics that characterized local and global divergences between the two datasets. All analyses were performed independently for the electrophysiology-only dataset (E→e-type) and the morpho-electrophysiological dataset (M+E→e-type), and for both augmentation levels (N = 5000 and N = 10 000 synthetic samples per class). To qualitatively evaluate the alignment between real and synthetic feature distributions, we generated merged kernel density estimates (KDEs) for all electrophysiological and morphological variables. For each feature, the empirical density of real training neurons, real held-out test neurons, and SMOTE-generated synthetic neurons was plotted together. This allowed visual inspection of (i) distributional overlap between synthetic and real data, (ii) agreement between synthetic samples and held-out biological variability, and (iii) deviations in multimodal or high-variance features. These merged density plots formed the basis of Figure 5. To quantify distributional similarity, we performed class-conditional Kolmogorov–Smirnov tests between real and synthetic neurons. For each neuronal class and each feature, the KS statistic and p-value were computed by comparing the empirical distributions of the real and synthetic samples. This produced a detailed map of univariate deviations across the entire feature set, highlighting both well-aligned and mismatched dimensions. We computed feature-wise MAE between synthetic neurons and real neurons for each class. MAE was estimated separately using the real training set and the held-out test set to distinguish fidelity of reconstruction from generalization to unseen biological variability. This produced two class-by-feature MAE matrices per dataset and per augmentation level, visualized as heatmaps (Figure 6). These matrices allowed systematic identification of which neuronal subclasses and which features were best reproduced or most challenging for synthetic generation. To quantify global divergence in high-dimensional phenotype space, we computed the Euclidean distance between the mean feature vector of each real class and the corresponding synthetic class mean. This measure captured how accurately synthetic neurons reconstructed the overall class signature, integrating information across all features simultaneously. To contextualize synthetic-to-real distances, we estimated the intrinsic biological diversity of each dataset by computing pairwise distances between the mean feature profiles of real neuronal classes. The resulting distribution of real inter-class distances served as a biological variability baseline. Synthetic-to-real distances were then statistically compared to this baseline using a Mann–Whitney U test (two-sided). This analysis determined whether synthetic neurons remained within the natural phenotypic envelope observed among real cell types or whether they exceeded biologically plausible limits (Marouf et al., 2020). All analyses were implemented in Python using pandas, numpy, scipy, seaborn, and matplotlib. 

\subsection{Classical Hardware}

All classical machine learning algorithms were executed on a workstation equipped with an AMD Ryzen 9 3900X 12-core (24-thread) processor, 32 GB of RAM, and an Nvidia GeForce RTX 3070 GPU (driver version 510.85.02, CUDA version 11.6). Deep generative models---including GANs, VAEs, Normalizing Flows, and Denoising Diffusion Probabilistic Models (DDPMs)---were trained and evaluated using Google Colab Pro cloud resources, which provided access to Nvidia GPUs (CUDA version 11.2 or higher).

\subsection{Code Availability}

All algorithms and analyses were implemented in Python 3 (https://www.python.org/). The classification and synthetic data generation scripts are available as open-source on GitHub at https://github.com/xaviervasques/NeuronalClassification. Classical machine learning pipelines and data manipulations were performed using scikit-learn (https://scikit-learn.org/stable/). All scripts, including those for data preprocessing, augmentation, classification, and evaluation, are provided in the GitHub repository to ensure full reproducibility. Neuron feature extraction relied on the MIES electrophysiology acquisition software (https://github.com/alleninstitute/mies), with morphological reconstructions processed using Vaa3D and the Mozak extension (http://www.vaa3d.org, https://github.com/Vaa3D). Electrophysiological and morphological feature analysis was performed with open-source tools from the Allen SDK (https://github.com/alleninstitute/allensdk) and IPFX (https://github.com/alleninstitute/ipfx) repositories. Clustering analyses leveraged tools from the Allen Institute’s DRCME repository (https://github.com/alleninstitute/drcme). All code, pipelines, and instructions to reproduce the full study, including synthetic data generation and classifier evaluation, are available in the project repository.

\section*{}

\end{document}